\documentclass[sn-mathphys-num,Numbered]{sn-jnl}

\usepackage{xurl}
\usepackage{graphicx}%
\usepackage{lmodern}%
\usepackage[export]{adjustbox}%
\usepackage{multirow}%
\usepackage{amsmath,amssymb,amsfonts}%
\usepackage{amsthm}%
\usepackage{mathrsfs}%
\usepackage[title]{appendix}%
\usepackage[dvipsnames]{xcolor}%
\usepackage{textcomp}%
\usepackage{manyfoot}%
\usepackage{booktabs}%
\usepackage{algorithm}%
\usepackage{algorithmicx}%
\usepackage{algpseudocode}%
\usepackage{listings}%
\usepackage{romannum}
\usepackage{nicefrac}
\usepackage{physics}
\usepackage{comment}

\usepackage{ulem}

\newcommand{\didv}{{d}\textit{I}/{d}\textit{V} }

\newcommand{\bSh}

\begin{document}

\title{Sensing a magnetic rare-earth surface alloy by proximity effect with an open-shell nanographene}

\author[1]{\fnm{Nicol\`{o}} \sur{Bassi}}
\author[2,3]{\fnm{Jan} \sur{Wilhelm}}
\author[1]{\fnm{Nils} \sur{Krane}}
\author[1]{\fnm{Feifei} \sur{Xiang}}
\author[4]{\fnm{Patrícia} \sur{\v{C}melov\'{a}}}
\author[1]{\fnm{Elia} \sur{Turco}}
\author[5]{\fnm{Pierluigi} \sur{Gargiani}}
\author[1]{\fnm{Carlo} \sur{A. Pignedoli}}
\author[4]{\fnm{Michal} \sur{Jurí\v{c}ek}}
\author[1,6]{\fnm{Roman} \sur{Fasel}}
\author[7]{\fnm{Richard} \sur{Korytár}}
\author*[1]{\fnm{Pascal} \sur{Ruffieux}}\email{pascal.ruffieux@empa.ch}

\affil[1]{Empa - Swiss Federal Laboratories for Materials Science and Technology, Dübendorf, Switzerland}

\affil[2]{Regensburg Center for Ultrafast Nanoscopy (RUN), University of Regensburg, 93053 Regensburg, Germany}

\affil[3]{Institute of Theoretical Physics, University of Regensburg, 93053 Regensburg, Germany}

\affil[4]{Department of Chemistry, University of Zurich, Winterthurerstrasse 190, 8057 Zurich, Switzerland}

\affil[5]{ALBA Synchrotron Light Source, E-08290 Cerdanyola del Vall\`es, Barcelona, Catalonia, Spain}

\affil[6]{Department of Chemistry, Biochemistry and Pharmaceutical Sciences, University of Bern, Bern, Switzerland}

\affil[7]{Department of Condensed Matter Physics, Faculty of Mathematics and Physics, Charles University, Ke Karlovu 5, 121 16, Praha 2, Czech Republic}

\keywords{Open-shell molecules, spin, phenalenyl,rare-earth surface, proximity interaction, scanning tunneling microscopy, XMCD}

\maketitle

\clearpage


\clearpage

\section{Abstract}\label{sec0}
Open-shell nanographenes have attracted significant attention due to their structurally tunable spin ground state. While most characterization has been conducted on weakly-interacting substrates such as noble metals, the influence of magnetic surfaces remains largely unexplored. In this study, we investigate how TbAu\textsubscript{2},  a rare-earth-element-based surface alloy,  affects the magnetic properties of phenalenyl (or [2]triangulene (2T)), the smallest spin-\nicefrac{1}{2} nanographene. Scanning tunneling spectroscopy (STS) measurements reveal a striking contrast: while 2T on Au(111) exhibits a zero-bias Kondo resonance - a hallmark of a spin-\nicefrac{1}{2} impurity screened by the conduction electrons of the underlying metal - deposition on TbAu\textsubscript{2} induces a symmetric splitting of this feature by approximately 20~mV. We attribute this splitting to a strong proximity-induced interaction with the ferromagnetic out-of-plane magnetization of TbAu\textsubscript{2}.  Moreover, our combined experimental and first-principles analysis demonstrates that this interaction is spatially modulated, following the periodicity of the TbAu\textsubscript{2} surface superstructure. These findings highlight that TbAu\textsubscript{2} serves as a viable platform for stabilizing and probing the magnetic properties of spin-\nicefrac{1}{2} nanographenes, opening new avenues for the integration of $\pi$-magnetic materials with magnetic substrates. 

\clearpage

\section{Introduction}\label{sec1}

Molecular magnets have emerged as promising candidates for spintronic applications due to their inherently tunable properties~\cite{pandey2023perspective,sanvito2010rise}. In contrast to conventional inorganic materials, molecular systems are expected to exhibit a longer decoherence time and offer greater flexibility in tailoring their spin properties through chemical modifications~\cite{bobbert2010makes,niu2016hyperfine,dediu2009spin}. Prominent examples for molecular spintronics include spin cross-over (SCO) molecules where the spin ground state can be modulated by external stimuli~\cite{chen2023observation,miyamachi2012robust,zhang2020anomalous,thakur2021thermal}. 
More recently, open-shell nanographenes with structurally controlled spin ground states have unlocked exciting possibilities for metal-free magnetism. Through on-surface synthesis, these molecular building blocks can be covalently assembled into spin chains, enabling precise control over their coupling strength~\cite{mishra2020topological,mishra2021observation,zhao2024tunable,turco2023observation}. 
For the successful integration of these structures into spintronic devices, establishing well-defined interfaces, the so-called spinterface~\cite{sanvito2010rise,cinchetti2017activating},  with ferromagnetic substrates is a key aspect. This interface serves to impose a spin reference by lifting the spin degeneracy of the molecular system while minimizing chemical interactions to preserve the intrinsic properties. Achieving this balance is crucial for harnessing the full potential of molecular magnets in the next-generation spintronic devices. 

\indent The early search for suitable ferromagnetic substrates focused on ultrathin layers or epitaxially grown monolayers of Fe, Ni or  Co~\cite{bode1997scanning,haag2016epitaxial,morgenstern2007cobalt,chen2023observation,Fu2012,Kawahara2010}. Recently, rare-earth-element-based surface alloys with noble metals have garnered increasing interest as a distinct class of atomically thin ferromagnetic films combining the magnetic properties of the rare earth element together with the catalytic effect from the noble metal. Since the early work by Ortega et al. on GdAu\textsubscript{2} in 2010~\cite{corso2010111}, various alloys have been explored, including GdAu\textsubscript{2}~\cite{correa2017self,bazarnik2019atomically}, HoAu\textsubscript{2}~\cite{que2020two,fernandez2020influence} and TbAu\textsubscript{2}~\cite{que2020two,que2020surface}. These surface alloys exhibit a common superstructure motif, characterized by a long-range hexagonal superstructure from the lattice mismatch between the alloy layer and the substrate. Despite their structural similarities, their electronic and magnetic properties differ significantly due to variations in the filling of the 4f-subshells. For instance, GdAu\textsubscript{2} exhibits ferromagentism with an in-plane easy-magnetization-axis~\cite{fernandez2010self,ormaza2016high}, whereas  HoAu\textsubscript{2} features an out-of-plane easy-axis~\cite{fernandez2020influence}. The well-ordered superstructure of these surfaces make them attractive platforms for investigating magnetic nanomaterials. So far, the focus has been mainly on metal clusters~\cite{fernandez2010self,cavallin2014magnetism}. However, interactions between rare-earth-element based surface alloys and organic molecules remain largely unexplored. To date, studies have been limited to a few examples, such as copper phtalocyanines (CuPcs)~\cite{castrillo2023tuning}, with only one recent study on entirely carbon-based materials~\cite{brede2023detecting}.

\indent In this work, we investigate the interaction between the surface alloy TbAu\textsubscript{2} and an open-shell nanographene. After successfully growing a high-quality TbAu\textsubscript{2} surface with the expected structure, X-ray magnetic circular dichroism (XMCD) measurements confirm its preferential out-of-plane ferromagnetic properties. As a model open-shell nanographene, we select phenalenyl (or [2]triangulene (2T)). On Au(111), low-energy scanning tunneling microscopy (STM) and spectroscopy (STS) reveal a pronounced Kondo resonance at zero bias, characteristic of a spin  \nicefrac{1}{2} system~\cite{turco2023observation}. However, when characterized on TbAu\textsubscript{2}, the Kondo resonance undergoes a symmetric splitting of roughly 20~mV relative to the Fermi energy $E_\mathrm F$. This splitting is significantly larger than in previous reports of Kondo impurities coupled to ferromagnets, considering the localized nature
of Tb f-shell~\cite{Pasupathy2004,Kawahara2010,Fu2012,Choi2016,Ternes2020,Garnier2020}.
By combining first-principles calculations and quantum many-body modeling, we attribute this effect to an effective Zeeman splitting induced by the proximity with the ferromagnetic surface.
Furthermore, we find that the extent of the splitting is highly sensitive to the adsorption site within the surface superstructure, as variations in the local arrangement of Tb atoms can modulate the interaction.
These findings establish TbAu\textsubscript{2} as a promising out-of-plane ferromagnetic platform for investigating open-shell nanographenes, offering new opportunities of $\pi$-magnetism in proximity of rare-earth-element based surfaces.

\section{Results and Discussion}\label{sec2}

\smallskip

\begin{figure}[!htbp]%
\centering
\includegraphics[width=0.8 \textwidth]{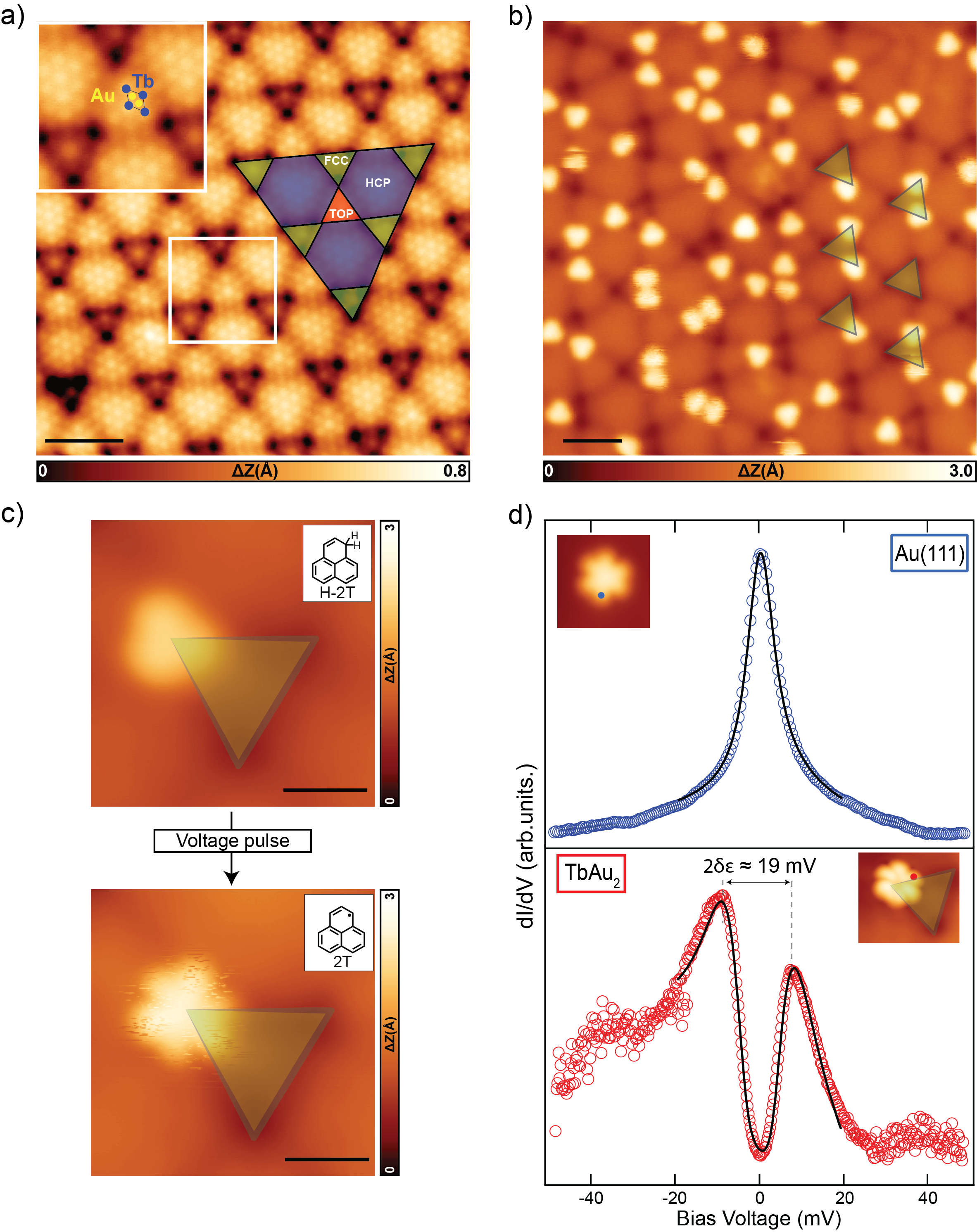}
\caption{\textbf{TbAu\textsubscript{2} and 2T molecules}. a) STM topography image of TbAu\textsubscript{2} which exhibits a periodic hexagonal superstructure. Three different regions are visible: dark-triangles, bright-triangles and hexagons (green, red and blue colors, respectively). Inset: High-resolution image resolving the Tb atoms. The TbAu\textsubscript{2} unit cell is outlined with Tb and Au atoms marked in blue and yellow. (scale bar: 4~nm, scanning parameters: -1~V, 200~pA). b) STM image of H-2T molecules on TbAu\textsubscript{2}. The green overlay marks the dark-triangle of the underlying surface superstructure (scale bar: 5~nm, scanning parameters: -1~V, 30~pA). c) High resolution STM image showing the change from H-2T (top) to 2T (bottom) after tip-manipulation. The shape of the molecule changes from a blunt triangle to a hexagonal structure. (scale bar: 500~pm; scanning parameters: -0.1~V, 20~pA) Insets: chemical structures of H-2T and 2T. d) \didv spectra of 2T on Au(111) (top panel in blue) and on TbAu\textsubscript{2} (bottom panel in red). On Au(111), the molecules exhibits a zero-bias peak with an half-width at half-maximum of 3~mV. Black curve: fit by Hurwitz-Fano \cite{turco2024demonstrating} formula. On TbAu\textsubscript{2}, there is a symmetric opening of $\sim$ 20~mV (open feedback parameters: $V$=50~mV, $I$=1~nA; V\textsubscript{rms}= 2~mV). Black curve: NRG fit of experimental \didv spectrum. } \label{Fig1}
\end{figure}

\indent The TbAu\textsubscript{2} surface alloy is prepared by sublimation of Tb atoms onto the Au(111) substrate kept at around 300~°C (see Methods section). As shown in Figure~\ref{Fig1} and in Figure~\ref{Fig SI1}, the resulting TbAu\textsubscript{2} layer exhibits a  well-ordered, high-quality hexagonal superstructure that extends over the entire surface with a periodicity of 3.64 $\pm$ 0.05~nm (confirmed by the sharp low-energy electron diffraction (LEED) pattern~\ref{Fig SI1}b). STM topography imaging shows that the surface is divided in three distinct domains appearing as dark triangles, bright triangles and hexagons separated by dislocation lines (Figure~\ref{Fig1}a). Similar strain-relief patterns are obtained for related rare-earth-element based alloys~\cite{corso2010111,fernandez2020influence,que2020two} as well as other metallic surface alloys~\cite{ruffieux2009mapping,ascrizzi2024dft,brune1994strain}. This superstructure is a consequence of the lattice mismatch between the Au(111) substrate (lattice constant of 2.97$\pm$ 0.025~\AA) and the TbAu\textsubscript{2} surface alloy (lattice constant 5.51$\pm$ 0.02~\AA). At this scanning range, Tb atoms appear as distinct dots as further illustrated by STM images acquired under different scanning parameters (see Figure~\ref{Fig SI1}).

\smallskip

The selected triangular phenalenyl (or 2T) molecule (chemical structure in Figure~\ref{Fig1}c and synthetic procedure reported in a previous work by Jurí\v{c}ek et al.~\cite{turco2023observation}) consists of an odd number of carbon atoms and, consequently, exhibits an unpaired $\pi$-electron (i.e. the molecule is in a spin  \nicefrac{1}{2} ground state). Due to the poor air-stability of open-shell nanographenes, the molecules sublimated are the non-reactive hydro-precursor phenalene (or H-2T). Figure~\ref{Fig1}b shows a low coverage STM image of H-2T molecules on TbAu\textsubscript{2}, revealing preferential adsorption at the corners of the dark triangular domains of the strain-relief pattern. This preferential adsorption, similarly to the behavior of H\textsubscript{2}-PC molecules on GdAu\textsubscript{2}~\cite{farnesi2019can}, comes from large atomic displacement at the dislocation lines, as discussed later. A high-resolution STM image of H-2T is reported in the top panel of Fig.~\ref{Fig1}c, where the molecule exhibits a blunt triangular shape. To obtain the target open-shell molecule, the extra H-atom is removed \textit{in-situ} by applying voltage pulses with the scanning probe sensor or by scanning at large bias voltages (around 3.0~V), following established protocols (see SI for a more detailed description)~\cite{wang2022aza, zhao2024tailoring}. As depicted in the bottom panel of Fig.~\ref{Fig1}c, the apparent STM topography drastically changes upon dehydrogenation, revealing six prominent lobes associated with the delocalized unpaired electron~\cite{turco2023observation}.

\smallskip

\indent On Au(111), the electronic properties of 2T near the Fermi energy $E_\mathrm F$ are characterized by a peak centered at 0~V, which has been identified as a Kondo resonance (Fig.~\ref{Fig1}d) with a half-width at half-maximum of 3 mV~\cite{turco2024demonstrating}. 
In contrast, on TbAu\textsubscript{2}, the \didv spectrum reveals a symmetric splitting of this resonance with respect to $E_\mathrm F$. As illustrated in Figure~\ref{Fig1}d, the gap present at the lobes shows a peak-to-peak splitting of $2\delta\varepsilon= 18~ \text{mV}$. Measurements on different molecules adsorbed at analogous locations within the strain-relief pattern yield a comparable gap ranging from 18 to 21~mV. An example of a complete activation procedure is provided in Figure~\ref{Fig SI2}.

\begin{figure}[!htbp]%
\centering
\includegraphics[width=0.5 \textwidth]{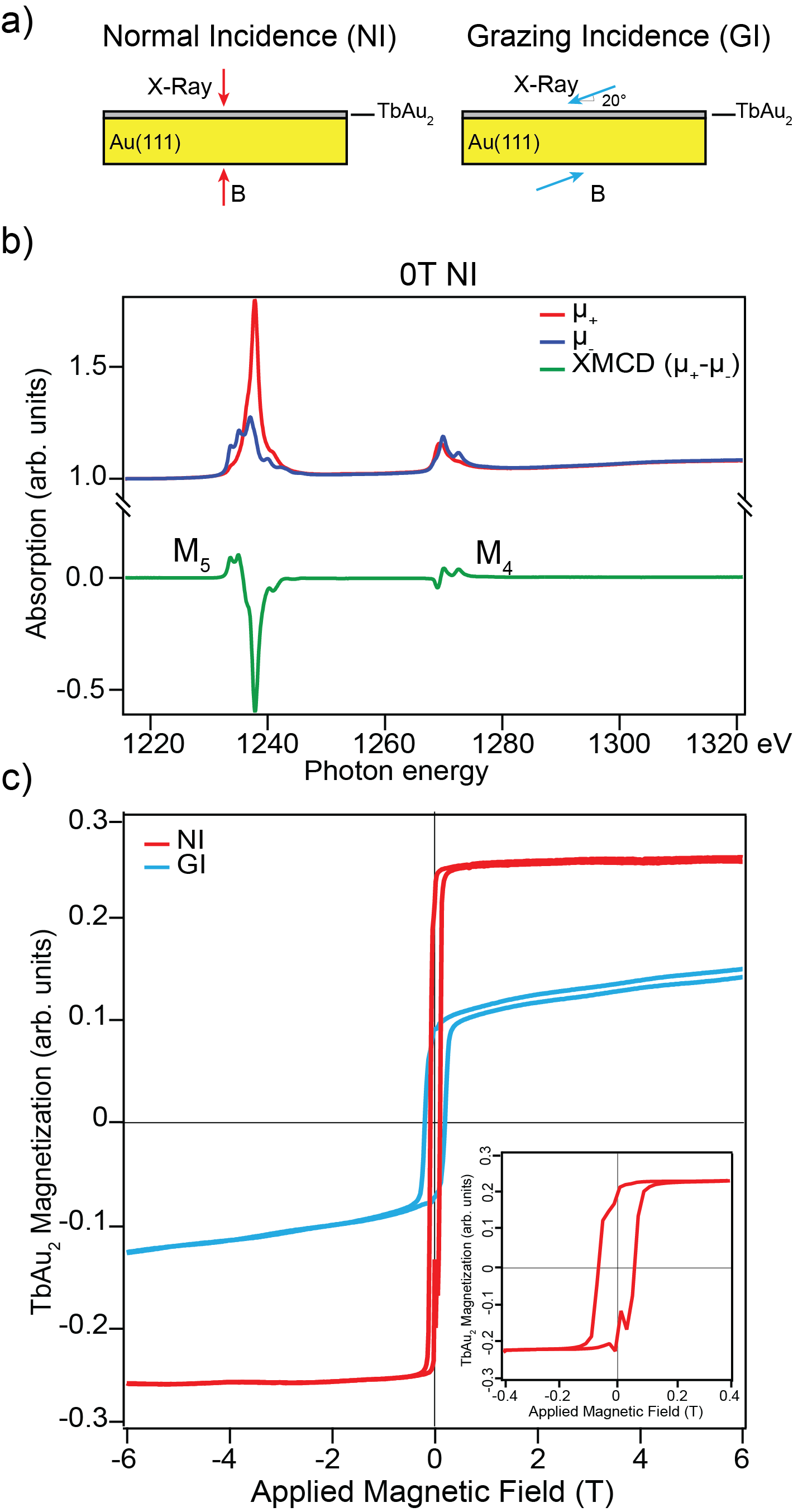}
\caption{\textbf{Magnetic properties of the TbAu\textsubscript{2}/Au(111) surface alloy.} a) Schematic representation of the two different experimental geometries: Normal Incidence (NI) and Grazing Incidence (GI, grazing angle 20~°). b) XAS spectra recorded at zero applied field in NI and T=3.5~K. The red and blue curves correspond to left-circularly polarized ($\mu_+$) and right-circularly polarized ($\mu_-$) light, respectively. The green curve represents the XMCD signal, which exhibits two prominent peaks at the Tb M\textsubscript{4,5} edges, confirming the ferromagnetic nature of the surface. c) Hysteresis loops measured at the Tb M\textsubscript{5} edge as a function of the applied magnetic field for two different configurations: NI (red) and GI (blue). Inset: A magnified view of the NI magnetization curve at low field, revealing a preferential out-of-plane easy-magnetization axis.} \label{Fig2}
\end{figure}

\smallskip 

To investigate the origin of the contrasting behavior of Au(111) and TbAu\textsubscript{2}, we conducted a more detailed characterization of the surface alloy. In a first step, the magnetic properties induced by the Tb atoms were probed by X-ray absorption spectroscopy (XAS) and X-ray magnetic circular dichroism (XMCD) (see Methods section). XAS and XMCD spectra (displayed in red, blue and green, respectively) acquired at a magnetic field of 0~T and at a sample temperature of $T=3.5$~K (Figure~\ref{Fig2}b) reveal that, after magnetic saturation of the sample, TbAu\textsubscript{2} still retains distinct magnetization signature, as evidenced by the prominent peak at the Tb M\textsubscript{5} edge (1238.4~eV), confirming its ferromagnetic nature (additional data for different configurations are presented in Figure~\ref{Fig SI3}). Hysteresis loops performed at the M\textsubscript{5} edge XMCD maximum (Figure~\ref{Fig2}c), in two different experimental configurations, namely, NI (normal incidence) or GI (grazing incidence) (schematically illustrated in Figure~\ref{Fig2}), confirm a preferential out-of-plane orientation of TbAu\textsubscript{2} magnetic axis. Specifically, the NI configuration (red curve) shows an almost immediate reversal and saturation following external field polarity, with the inset in Figure~\ref{Fig2}c highlighting hysteresis cycles at low field and a coercive field of approximately 50~mT. Contrary, the GI configuration (blue) grows linearly without reaching saturation in the available field range. Finally, by performing hysteresis measurements at progressively higher temperatures (see Figure~\ref{Fig SI3}) and applying Arrott plot analysis, we determine the Curie temperature $T\textsubscript{C}$ of the TbAu\textsubscript{2}/Au(111) surface alloy to be 13.5~K.

\smallskip

\begin{figure}[!htbp]%
\centering
\includegraphics[width= 0.8\textwidth]{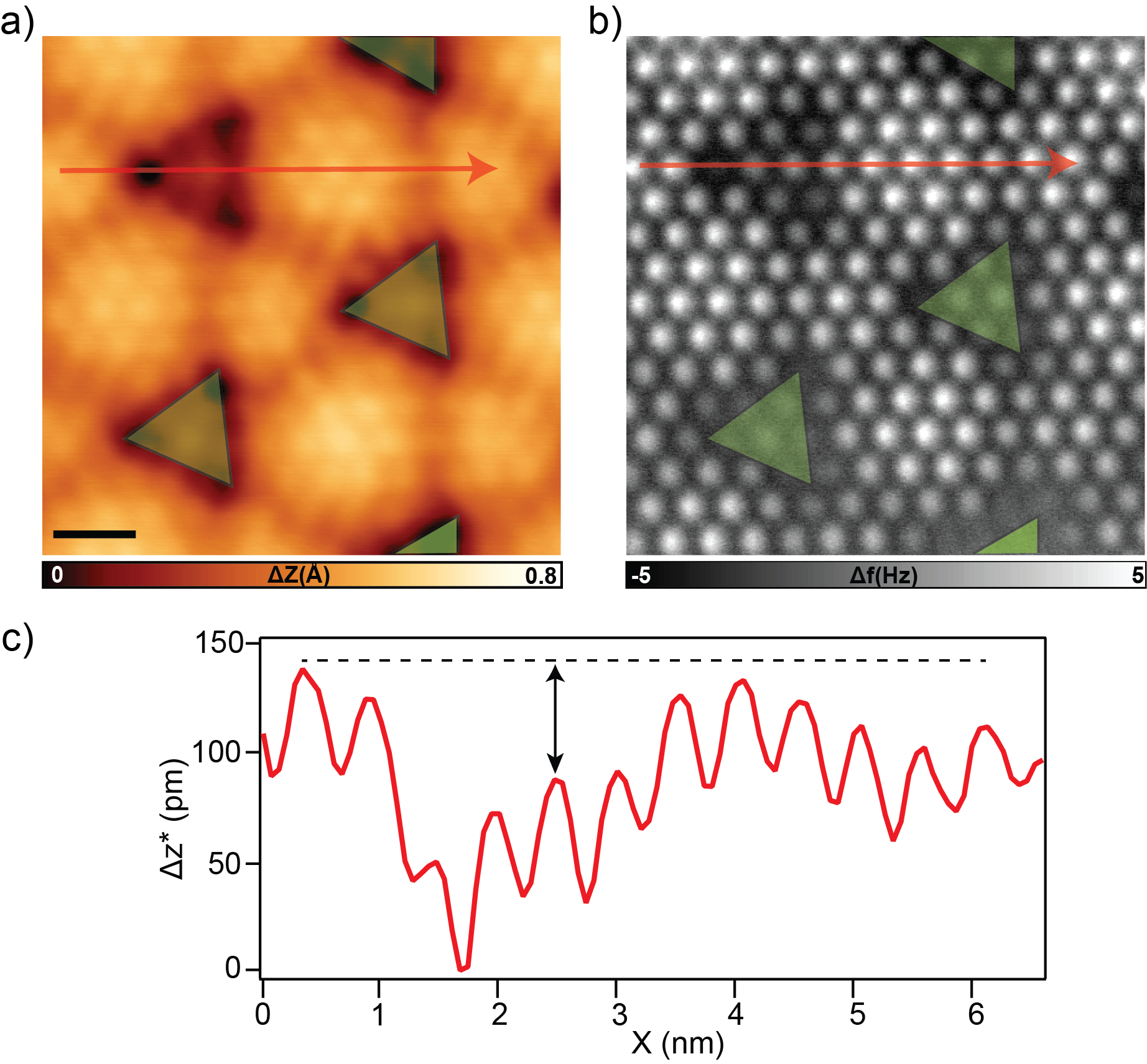}
\caption{\textbf{Structural corrugation of TbAu\textsubscript{2} on Au(111)}. a) STM topographic image of TbAu\textsubscript{2}. b) Corresponding AFM image, revealing height variations of Tb atoms along the surface pattern. The contrast indicates that the Tb atoms in the dark-triangle regions are positioned lower than those in the surrounding areas.  (scale bar: 1~nm; scanning parameters: -0.2~V, 200~pA. Open feedback parameters: $V$ = -0.2~V, $I$ = 500~pA, $\Delta z$ = -150~pm). c) Relative height $\Delta z^\ast$ extracted from frequency shift minima, reflecting the vertical displacement of Tb atoms along the strain-relief pattern. The data, acquired along the red line in the STM image (left), show a height variation of approximately 40~pm between the bright regions and the dark-triangle areas.} \label{Fig3}
\end{figure}

\indent The other difference between TbAu\textsubscript{2} and Au(111) is the presence of a periodic superstructure at the surface, arising from variations in vertical displacements of the atoms. 
We confirmed this height variation through non-contact atomic force microscopy (nc-AFM) measurements using a CO-functionalized tip. Constant-height $\Delta f$ images (Figure~\ref{Fig3}b) show a periodic structure with a lattice constant of 5.4 $\pm$ 0.2~\AA, matching that of TbAu\textsubscript{2}. Consequently, the structures observed in the AFM image are attributed to the Tb atoms in the surface layer, while the Au atoms in-between Tb are not visible.
In addition, a general contrast in the height of the Tb atoms is evident matching the superstructure seen in the STM images.
Since the AFM measurements were taken in the repulsive regime, a more positive (brighter) frequency shift corresponds to a topographically higher atom, whereas, the dark triangular regions in the STM images correspond to areas of topographically lower Tb atoms in the AFM.
To quantitatively estimate the relative heights of the Tb atoms, 100 $\Delta f(z)$ spectra were acquired along a line crossing the strain-relief pattern (red arrow in Figure~\ref{Fig3}a,b) and the relative height $\Delta z^\ast$ of the frequency shift minimum for each spectrum was extracted by a fit with a Lennard-Jones potential (see~\ref{Fig SI4} for more details)~\cite{schuler2013adsorption}.
The resulting topographic profile, shown in Figure~\ref{Fig3}c, reveals a relative height difference of about 40~pm between the bright areas in STM and the dark triangles. The largest height difference (60~pm) is reached at the corner of the triangles, where dislocation lines intersect.

\smallskip

This periodic vertical displacement is also reproduced in DFT simulations of TbAu\textsubscript{2} on Au(111) after relaxation (Figure~\ref{Fig SI5} and Methods for the computational details). In the relaxed geometry, a hexagonal superstructure is visible, with Tb and Au atoms (in blue and yellow colors, respectively) having periodically repeating height differences. This periodic pattern is corroborated by simulated STM images, which display a strain-relief pattern analogous to that observed experimentally. We further identified the displacements for two different stacking configurations: Top (green) and Hollow (red) (See panel SI~\ref{Fig SI5}b). In the Top configuration, where a Tb atom is positioned directly above an Au atom, the Tb atoms have a downward displacement of roughly 0.82~\AA~ relatively to the Au atoms, which will move downwards the Au atom directly underneath. Conversely, in the Hollow configuration, where a Tb atom is located at the interstitial site between three Au atoms, the Tb atoms are moved upwards by roughly 0.1~\AA~relative to the surrounding Au atoms.  

\begin{figure}[!htbp]%
\centering
\includegraphics[width=1.0 \textwidth]{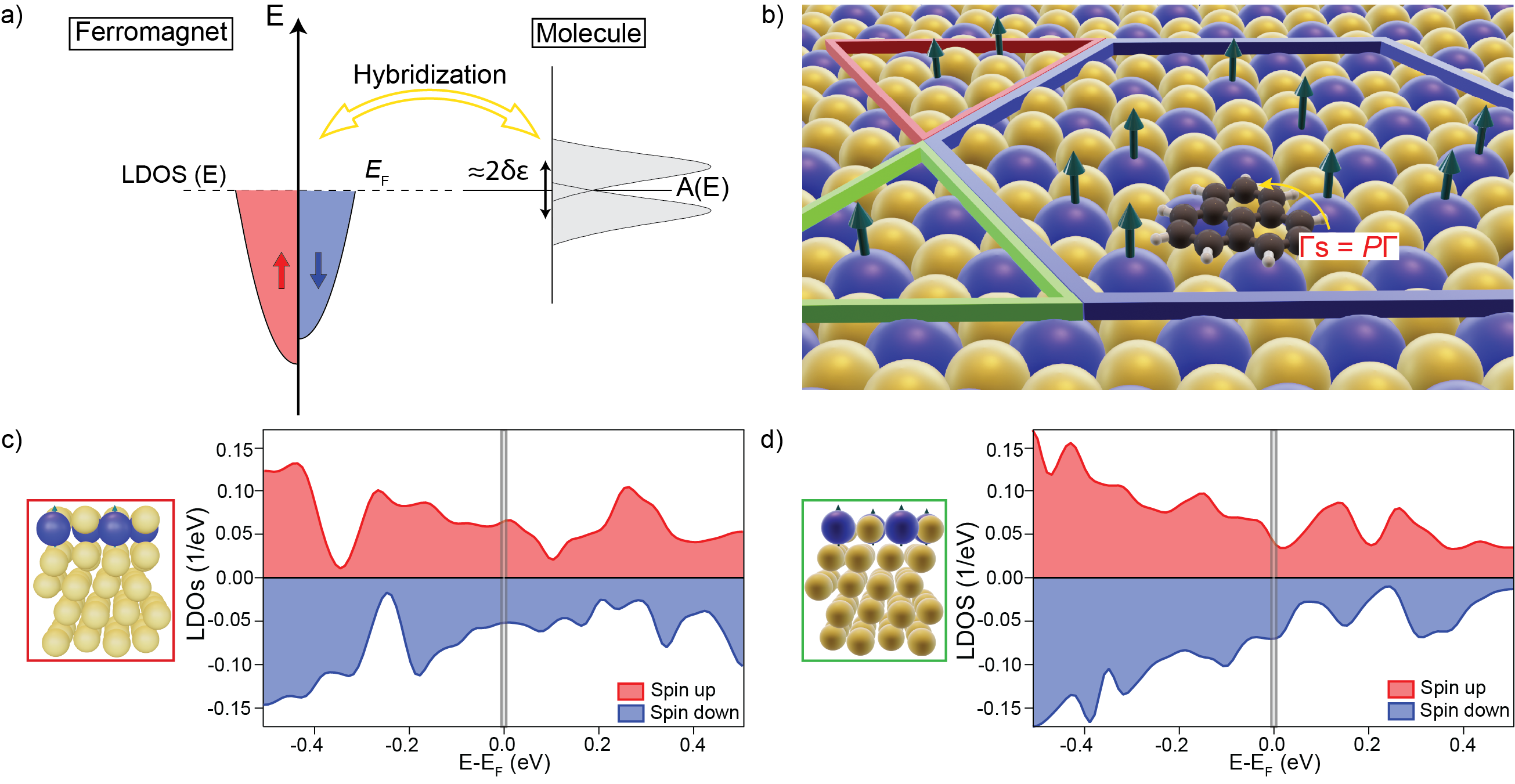}
\caption{\textbf{TbAu\textsubscript{2} interaction model}. a) Schematic representation of the interaction between TbAu\textsubscript{2} surface (left) and 2T molecule (right). The spin polarization of the TbAu\textsubscript{2} surface is modeled by different local density of states (LDOS) at the Fermi level ($E_\mathrm{F}$), with spin-down states represented in blue and spin-up states in red. This spin polarization induces a splitting in the spectral function of the 2T molecule at $E_\mathrm{F}$. b) Graphical representation of 2T molecules on TbAu\textsubscript{2}, with the magnetic properties of the surface modeled by out-of-plane arrows on the Tb atoms. c) Spin-polarized LDOS for the Top configuration of the TbAu\textsubscript{2} surface (the model for this geometry is shown on the left). d) Spin-polarized LDOS for the Hollow configuration, with the corresponding model on the left side. The two geometries show different LDOS, attributed to the different relative heights of the Tb atoms in the two configurations. In gray, the energy interval near $E_\mathrm{F}$ used in Eq. (2) is highlighted.} \label{Fig4}
\end{figure}

\smallskip

We now revisit our experimental observation of a split Kondo resonance in \didv spectra (Figure~\ref{Fig1}d).
To gain a better understanding of the mechanism responsible for the symmetric gap opening at $E_\mathrm F$, we employ the single-impurity Anderson model (SIAM)~\cite{anderson1961localized, krishna1980renormalization}. 
This model accounts for the hybridization between surface conduction bands and the singly occupied molecular orbital (SOMO) of the spin-\nicefrac{1}{2} molecule (see Methods for details).
In addition, the SIAM includes the on-site Coulomb repulsion $U$ when two electrons simultaneously occupy the SOMO. 
Due to the localized nature of Tb's 4\textit{f}-subshells, their effect on the adsorbed molecule is indirect. As observed in other rare-earth surface alloys~\cite{bazarnik2019atomically}, the interaction among Tb atoms is mediated by conduction electrons of the host metals via a RKKY mechanism. Thus, ferromagnetism is transmitted to the molecules through hybridization and not by direct exchange.
We capture this ferromagnetic proximity effect
as an effective Zeeman splitting on the SOMO denoted by $\delta\varepsilon$ (further justification is provided below). We calculate the  spectral function of the SIAM  using the numerical renormalization group  (NRG) (see Methods for further details)~\cite{krishna1980renormalization,vzitko2009energy,zitkoNRGLjubljana2021}.
For the  experimental \didv curve shown in Figure~\ref{Fig1}d,  we determine the optimal value of $\delta\varepsilon$  by minimizing the least-square deviation between the calculated density of states and the experimental data over the energy range [--\,20~mV, 20~mV]. 
This fitting procedure yields $\delta\varepsilon=9.4\,\text{~mV}$ (black continuous line in Figure~\ref{Fig1}d).
We conclude that the effective Zeeman splitting of the SOMO, resulting from ferromagnetic proximity, is in the order of 10~mV. Thus, the interaction with the ferromagnetic substrate removes the spin degeneracy and imprints a preferential out-of-plane orientation to the spin of the 2T molecule. 

\smallskip

The Anderson impurity in contact with one or more ferromagnetic leads has been studied extensively using many-body techniques
~\cite{martinek2003kondo,martinek2005gate,sindel2007kondo}.  
It has been demonstrated that the influence of the ferromagnet on the impurity is spectrally equivalent to the normal-metal problem with an effective Zeeman splitting~$\delta\varepsilon$ on the impurity, which justifies our simplified Hamiltonian.
Furthermore, the splitting is related to the variables of the ferromagnetic system by:
\begin{align}
\delta\varepsilon \simeq 
\frac{1}{\pi}\,\Gamma P,\;
\end{align}
\noindent
where $\Gamma$ is the SOMO broadening and  
\begin{align}
\hspace{1em}P=\frac{\rho_\text{F}^\uparrow - \rho_\text{F}^\downarrow }{\rho_\text{F}^\uparrow +\rho_\text{F}^\downarrow}
\end{align}
\noindent
is the spin polarization of the metal surface evaluated at $E_\mathrm F$ ($\rho^{\uparrow/\downarrow}_\text{F}$ are the electronic densities of states of spin up and down, respectively, of the metal surface). A graphical representation of the interactions is shown in Figure~\ref{Fig4}a, where the spin polarization of TbAu\textsubscript{2} is illustrated as a difference in spin population near $E_\mathrm F$. A schematic illustration is provided in Figure~\ref{Fig4}b.

\smallskip

In our case, TbAu\textsubscript{2} is a ferromagnetic surface, as confirmed by XMCD measurements, whose  spin polarization originates from the Tb atoms which hybridizes with the \textit{sp}-orbitals of the Au atoms in the alloy. Due to the periodic but distinct displacements of the atoms along the strain relief pattern, the density of states, and consequently the spin polarization $P$, is expected to vary accordingly. To confirm this, we evaluated the density of states at two different stacking geometries of Tb atoms with respect to the Au underneath (Hollow and Top) and extracted the values at $E_\mathrm F$ (Figure~\ref{Fig4}c and~\ref{Fig4}d show the local density of states (LDOS) from these two). As anticipated, the spin polarization $P$ differs between these two cases, with the Hollow geometry exhibiting a larger value compared to the Top (0.14 vs 0.09). Using these values in Eq.~1 and assuming $\Gamma = 300$~meV, a typical coupling value for organic molecules~\cite{mugarza2012electronic,soe2009direct,zuzak2017nonacene},  we expect the Zeeman splitting $\delta\varepsilon$ to vary between 10.5~mV in the Hollow geometry and 6.9~mV in the Top geometry.

\smallskip

\indent To verify the position dependence of the energy splitting, we deposited a high coverage of H-2T molecules to uniformly cover the surface (see Figure~\ref{Fig SI6} and~\ref{Fig SI7} for images of progressive coverage increases).
\begin{figure}[!htbp]%
\centering
\includegraphics[width=\textwidth]{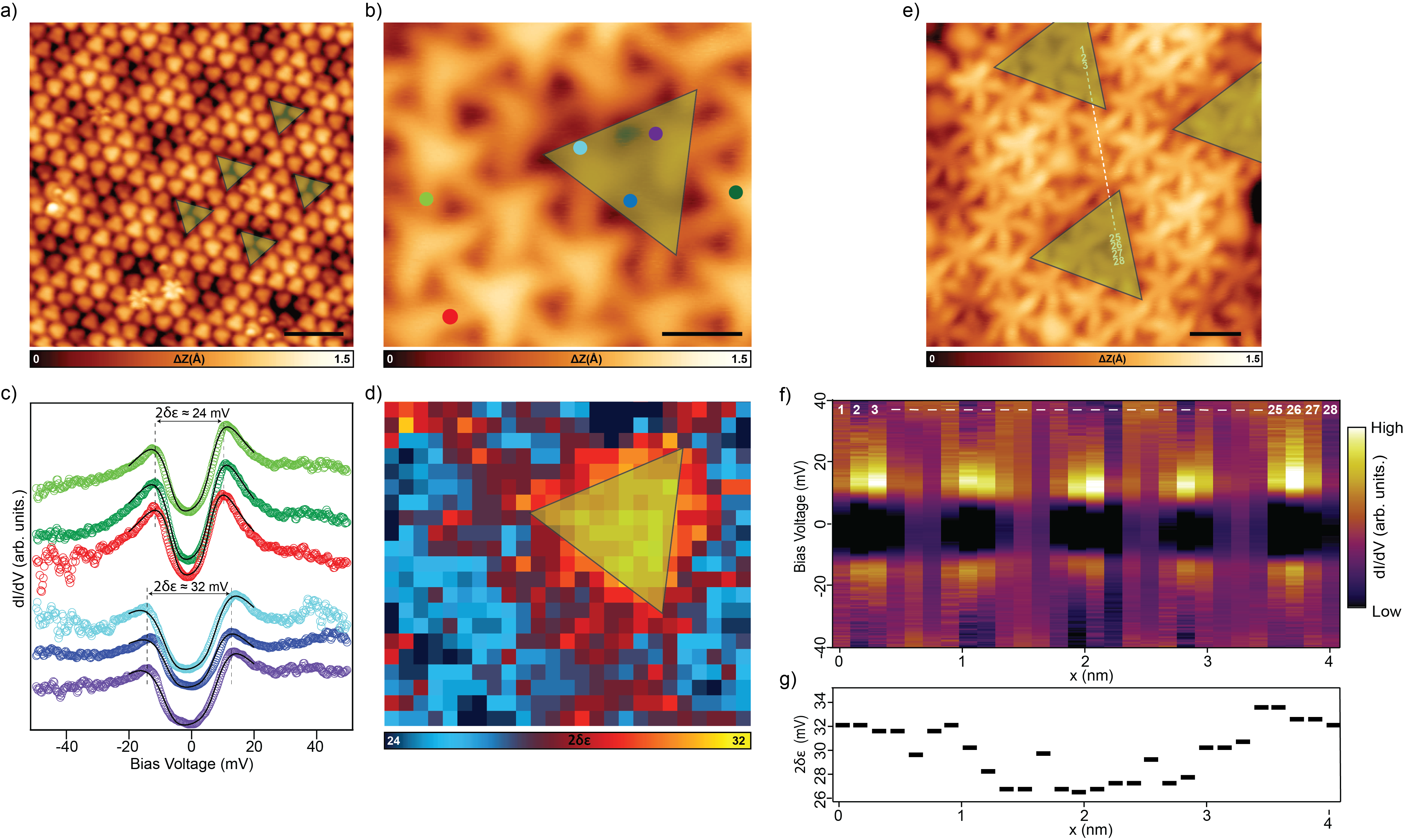}
\caption{\textbf{Monolayer of 2T on TbAu\textsubscript{2}}. a) STM image of TbAu\textsubscript{2} after deposition of a monolayer of H-2T. The molecules are densely packed and follow the surface structure underneath (scale bar: 2~nm, scanning parameters: -0.15~V, 100~pA). b) STM image of a small region of the surface after tip-induced manipulation of the H-2T molecules (scale bar: 1~nm, scanning parameters: -0.1~V, 100~pA). c) \didv spectra on different points on the surface (see panel b). Spectra taken from different molecules are show a splitting of the peak, with the gap ranging from 22~mV to 28~mV at the dark-triangle regions.  (Open feedback parameters: $V$: -50~mV, $I$~1 nA; V\textsubscript{rms}: 1~mV). d) Map of the effective Zeeman splitting~$2\delta\varepsilon$  of the same region in panel b. $2\delta\varepsilon$ is extracted from the fit of NRG spectral functions to the \didv curves. The largest splitting (yellow) is observed in the Hollow regions. e) STM topographic image of TbAu\textsubscript{2} and 2T molecules on top (scale bar: 1~nm, scanning parameters: -0.04~V, 100~pA) f) 3D plot of spectra taken along a line (white dotted line in panel e). A periodic pattern is observed in the intensity, with larger splitting visible at the ends of the line. g) Plot showing the different $2\delta\varepsilon$ values extracted from the spectra in panel (f). The highest values are observed at the ends of the line, with a distance between the peaks corresponding to the periodicity of the strain-relief pattern.} \label{Fig5}
\end{figure}
Upon reaching full monolayer coverage of H-2T molecules on TbAu\textsubscript{2}, an ordered pattern is attained (Figure~\ref{Fig5}a). The molecules are densely packed and the underlying surface superstructure is still visible, with the periodicity of the dark triangles (green) matching that of the strain relief pattern. Due to the high density of molecules, manipulation of H-2T was carried out in multiple steps (see SI). STM image of a region after complete tip-induced manipulation is shown in panel~\ref{Fig5}b, where the underlying superstructure is still evident. Low-bias spectroscopy taken on different positions reveals varying gaps (Figure~\ref{Fig5}c). Spectroscopy curves measured at the Hollow regions (blue-like curves in Figure~\ref{Fig5}c) show larger splitting compared to the one measured outside (32 and 24~mV, respectively). 
To map the effective Zeeman splitting, we measured a dense grid of spectroscopy curves in the same region shown in Fig.~\ref{Fig5}b. For a quantitative evaluation of the splitting, we again solve the SIAM with the effective Zeeman splitting, as previously done for \didv curve in Figure \ref{Fig1}d, and extracted the splitting~$\delta\varepsilon$.
The resulting coupling map is reported in panel~\ref{Fig5}d. This plot confirms the varying splitting across the surface, with a progressive change in the $2\delta\varepsilon$ values, going from approximately 24~mV to 32~mV. This variation matches the expectations from the previous modeling, with the Hollow configuration (green triangles) showing the higher splitting values.

\smallskip

\indent  To further confirm the correlation with the strain-relief pattern, we examined whether the spatial variation of~$\delta\varepsilon$ matches that of the TbAu\textsubscript{2} surface superstructure. A series of \didv spectra was acquired along a line bridging two different Hollow regions (white dashed line in Figure~\ref{Fig5}e). An intensity plot summarizing the curves is shown in Figure~\ref{Fig5}f (see Figure~\ref{Fig SI9} for individual spectra). The spectra taken at the extremes of the line (i.e on dark-triangles or FCC regions) exhibit a larger zero-bias splitting than those in the middle. The distance between these two units is 3.38~nm, the same than the periodicity of the surface superstructure. This position dependence is also evident after extracting the effective Zeeman splitting~$\delta\epsilon$ (Figure~\ref{Fig5}f a plot summarizing the values). The highest values are observed at the extremes of the line, while lower ones are reported in the central part. Finally, we also mapped the effective Zeeman splitting across a large surface area (Figure~\ref{Fig SI10}a). Spectroscopy curves taken from dark triangles (blue-like spectra) once again show a larger gap compared to those measured outside these regions.  A similar trend is observed when measuring a grid of spectra in the same area, where the splitting is larger in the dark triangles, and the periodicity aligns with the strain-relief pattern of TbAu\textsubscript{2}/Au(111).
\backmatter

\section{Conclusions}\label{conc1}

In conclusion, we have characterized how the magnetic properties of an open-shell nanographene are modified when interacting with a ferromagnetic surface. The selected substrate, TbAu\textsubscript{2}, a rare-earth-element based surface alloy, was confirmed to exhibit a high-quality periodic superstructure by STM techniques. Through XAS and XMCD technique we verified its magnetic nature, identifying it as a ferromagnetic surface with an out-of-plane easy axis. 
For the open-shell spin-\nicefrac{1}{2} nanographene, we selected [2]triangulene (2T). The electronic properties near the Fermi energy, characterized by STS, reflect the opening of a symmetric gap of approximately 20~mV, confined at the lobes of the molecule.  We attribute the observed gap to the proximity induced magnetization with the TbAu\textsubscript{2} substrate and modeled the interaction as an effective Zeeman splitting $\delta\epsilon$. This interpretation was further supported by fitting the experimental \didv curve with the spectral function of a corresponding single-impurity Anderson model.
Additionally, due to the periodic displacement of the Tb atoms along the surface, we expect variation in this interaction. This was confirmed both by first-principles calculations, evaluating the spin polarization of the LDOS, and experimentally, by measuring changes in the gap opening at different positions of the surface superstructure.
The results reported herein demonstrate that the magnetic properties of an open-shell nanographene are preserved on TbAu\textsubscript{2}, and that the ferromagnetic surface lifts the spin degeneracy giving a preferential orientation to the molecular spin.

\section{Methods}\label{method}

Au(111) single crystal surfaces (MaTeck GmbH) were prepared by iterative Ar\textsuperscript{+} sputtering and annealing cycles. Prior to sublimation of other species, the surface structure and cleanliness was checked by STM imaging. Tb atoms were deposited onto clean Au(111) by a commercial e-beam evaporator (EFM-3T purchased from Omicron) by a Tb rod (Goodfellow, purity 99\% diameter: 2~mm, length: 5~cm). The Au(111) crystal was kept at roughly 330~°C during the deposition and the process was followed by a post annealing treatment at the same temperature. By adjusting the coverage, different density of grain boundaries could be achieved. 

The molecular precursors phenalene (H-2T) were synthesized by solution chemistry following already reported procedure~\cite{turco2023observation}. The H-2T molecular powder were deposited from a home-built Knudsen cell evaporator at 130~°C onto the crystal surfaces held at room temperature. 

\indent STM and AFM measurements were performed with a commercial low-temperature (4.5~K) STM/AFM from Scienta Omicron, equipped with a qPlus tuning fork sensor~\cite{Giessibl2000} and operated at base pressure below 5×10\textsuperscript{-11}~mbar. STM images were acquired in constant-current mode (overview and high-resolution imaging), \didv spectra were acquired in constant-height mode, the feedback-loop off parameters are reported in the captions. Differential conductance \didv spectra were obtained with a lock-in amplifier. Bond-resolved nc-AFM images and $\Delta f(z)$ curves were acquired in constant-height mode with CO-functionalized tips at low bias voltages. Open feedback parameters and lowering of the tip height $\Delta z$ are reported in the Figure caption. The data were processed with Wavemetrics Igor Pro software.

\indent XMCD experiments were realized at the BOREAS beamline of ALBA synchrotron, Spain~\cite{Barla2016a}. The samples were prepared \textit{in-situ} and the quality checked by LEED. Absorption spectra were acquired in total electron yield mode measuring the sample drain current at two different geometry: out-of plane (normal incidence of the beam) and in-plane (20° incidence of the beam). The magnetic field was collinear to the light propagation. The measurements were performed between 2~K and 20~K with a variable magnetic field between $\pm$ 6~T.

\smallskip

\subsection{Single-impurity Anderson model with ferromagnetic substrate and fitting procedure}
To better understand the mechanism leading to the split peak in \didv spectra, we employ the single-impurity Anderson model (SIAM)~\cite{anderson1961localized}
that couples the singly occupied molecular orbital (SOMO) of 2T to the conduction bands of the surface:
\begin{align}
\label{eq:siam}
\hat{H}_\text{SIAM} =   
\hat{H}_\text{SOMO} +
\hat{H}_\text{band} +
\hat{H}_\text{coup}\,.
\end{align}
$\hat{H}_\text{SOMO}$ describes the SOMO of 2T,
\begin{align}
\hat{H}_\text{SOMO} = \sum_{\sigma=\uparrow,\downarrow}\varepsilon_\text{SOMO}\,\hat{n}_{\text{SOMO},\sigma} + 
U\, \hat{n}_{\text{SOMO},\uparrow} \,\hat{n}_{\text{SOMO},\downarrow}   \,, 
\end{align}
where $\hat{n}_{\text{SOMO},\sigma}$ is the number operator counting the number of electrons of spin~$\sigma$ in the SOMO,  $U$ is the on-site Coulomb repulsion in case two electrons with opposite spin are occupying the SOMO and we set the SOMO orbital energy $\varepsilon_\text{SOMO}=-U/2$~\cite{zonda2021resolving}.  
$\hat{H}_\text{band}$ is the Hamiltonian of the surface conduction bands~$k$, 
\begin{align}
\label{eq:band}
\hat{H}_\text{band} = \sum_{k\sigma}\varepsilon_{k\sigma}\,\hat{n}_{k\sigma}\,
\,, 
\end{align}
where $\varepsilon_{k\sigma}$ is the spin-dependent
band structure and $\hat{n}_{k\sigma}=\hat{c}^\dagger_{k\sigma}\hat{c}_{k\sigma} $ is the number operator, $\hat{c}^\dagger_{k\sigma}$ the creation operator and $\hat{c}_{k\sigma}$ the annihilation operator of an electron in state $k\sigma$. 
$\hat{H}_\text{coup}$ is the Hamiltonian coupling the impurity level and the surface conduction bands,
\begin{align}
\label{eq:ht}
\hat{H}_\text{coup} = \sum_{k\sigma}
t_k\,\hat{c}^\dagger_{\text{SOMO},\sigma}\hat{c}_{k\sigma}+
t_k^*\,\hat{c}^\dagger_{k\sigma}\hat{c}_{\text{SOMO},\sigma}
\,.
\end{align}

In writing the Eqs.~(\ref{eq:siam}-\ref{eq:ht}) we assumed that ferromagnetism is expressed through
the spin-dependent $\varepsilon_{k\sigma}$ only.
Because the open f-shells of Tb are very localized, they influence the adsorbates only indirectly
through the hybridization with sp-orbitals.
Therefore, spin-splitting penetrates to the molecule by ferromagnetic proximity and not by a
direct exchange.
The influence of the conduction band on the SOMO is entirely contained in the hybridization function
\[
\Gamma_\sigma( E ) = \pi\sum_k |t_k|^2\delta(E - \varepsilon_{k\sigma}),
\]
which represents the spin-dependent broadening of the SOMO level.
The model (\ref{eq:siam}) endowed with the spin-dependent hybridization was studied extensively
in~\cite{martinek2003kondo, martinek2005gate, sindel2007kondo}. It was found that the
effects of the ferromagnetic host on the observables of the impurity are to a large
degree controlled by the parameter 
\[
\Gamma_\text{S} := \Gamma_\uparrow - \Gamma_\downarrow
\]
where $\Gamma_\sigma$ is $\Gamma_\sigma(E)$ evaluated at the Fermi energy. Under a perturbative renormalization scaling,
the spin-dependent hybridization induces an on-site splitting term
\begin{equation}
\label{eq:hx}
\hat H_\mathrm X = \delta\varepsilon (\hat{n}_{\text{SOMO},\uparrow} -\hat{n}_{\text{SOMO},\downarrow}).
\end{equation}
For the sake of illustration, we recollect an explicit expression for a flat band~\cite{martinek2003kondo, martinek2005gate, sindel2007kondo}
\begin{equation}
\label{eq:exchf}
\delta\varepsilon  = \frac 1{\pi} \Gamma_\text{S}
\log\left|\frac {
\varepsilon_\text{SOMO}
}{U + 
\varepsilon_\text{SOMO}
}\right|.
\end{equation}
The expression \eqref{eq:hx} can be understood as an effective Zeeman
field felt by the spin of the SOMO.
It was demonstrated in the Fig.~2 of Ref.~\cite{martinek2003kondo} that the differential
conductance of the impurity shows a pair of peaks near zero bias at voltages approximately
$\pm \delta\varepsilon$.

In this work we want to quantify the effect of the ferromagnetic TbAu\textsubscript{2} host on the SOMO by fitting
experimental differential conductance curves with numerical spectral functions for the SIAM with
a least-squares procedure.
As shown above, $\Gamma_\text{S}$ is such a quantifier. However, the fitted value of
$\Gamma_\text{S}$ would strongly correlate
with other parameters of the SIAM, namely,
$U$ and $\varepsilon_0$ (evident from Eq.~(\ref{eq:exchf})) and $\Gamma_\uparrow + \Gamma_\downarrow$.
The measured \didv do not extend to sufficiently high biases in order
to obtain all parameters of the SIAM reliably. Therefore, we resort to a pragmatic
approach by (i) fixing some parameters of the SIAM to physically meaningful values
and (ii) mimicking the ferromagnetic proximity by introducing an effective 
exchange field of the form (\ref{eq:exchf}) with $ \delta\varepsilon$
as a fitting parameter and setting $\Gamma_\uparrow = \Gamma_\downarrow$. The substitution
of the ferromagnetic host by an effective exchange field
was also supported by the conclusions of the combined theoretical - experimental work~\cite{gaass2011universality}.

As an alternative way to estimate $\delta\varepsilon$ directly from Eq.~\eqref{eq:hx}, we estimate $\Gamma_\text{S}$~\cite{sindel2007kondo} 
\begin{align}
    \Gamma_\text{s} = \Gamma P\;,
    \hspace{2em} P=
    \frac{\rho^\uparrow_\text{F} - \rho^\downarrow_\text{F} }{
\rho^\uparrow_\text{F} +\rho^\downarrow_\text{F} 
    }
\end{align}
where $P$ is the spin polarization of the metal surface,  $\rho^{\uparrow/\downarrow}_\text{F}$ is the density of states of $\uparrow/\downarrow$ electrons at the Fermi level of the metal surface and $\Gamma $ is a typical coupling of an organic molecule to a metal substrate. 
We estimate $P=0.05$ from ab-initio simulations,  $\Gamma = 300$\,meV and $\log | 
 \varepsilon_\text{SOMO}/(U + 
\varepsilon_\text{SOMO})|\simeq1$ which leads to $\delta\varepsilon =5$\,meV,
consistent with the value obtained by fitting (see below).

 The final Hamiltonian of our model is 
 \begin{align}
     \hat{H} = \hat{H}_\text{SIAM} + \hat{H}_\text{exch},
 \end{align}
where $\hat{H}_\text{exch}$ has the same form as \eqref{eq:exchf} with $\delta\varepsilon$ being
a free fitting parameter.

We use the numerical renormalization group (NRG)~\cite{krishna1980renormalization,vzitko2009energy},  to compute the spectral function of the SIAM with the ``Zeeman field''
$\delta\varepsilon$.
We assume energy and spin-independent hybridization functions 
$\Gamma(\varepsilon) \equiv \Gamma\,\theta(D^2- \varepsilon^2), D=2U$~\cite{zonda2021resolving} and an electronic temperature of $T=4.3$\,K.
Using NRG, we compute the spectral function~$\rho[U,\delta\varepsilon,\Gamma](\varepsilon)$
as function of energy~$\varepsilon$ for fixed~$U=0.5$\,eV and a large set of $\Gamma$,$\delta\varepsilon$.  
We then fit the experimental \didv curves by
\begin{align}
  \didv(V) \simeq 
 a\, \rho[U,\delta\varepsilon,\Gamma](eV) 
  +
  b\,V + c
\end{align}
where $V$ is the bias voltage and $a,b,c$ are fit parameters. 
The term $b\,V+c$ accounts for an increasing background density of states. 

\subsection{Computational Methods}
The geometry optimization of the TbAu\textsubscript{2} surface and the STM simulation have been performd with an AiiDAlab~\cite{yakutovich2021aiidalab} application based on  AiiDA~\cite{pizzi2016aiida} workflows for the DFT code CP2K~\cite{hutter2014cp2k}
The surface was modeled in the repeated slab scheme. The simulation cell consisted of four atomic layers of Au along the [111] direction and a layer of TbAu\textsubscript{2}. A layer of hydrogen atoms was used to passivate one side of the slab to suppress the Au(111) surface state. A vacuum of 30 \AA~was included in the simulation cell to decouple the system from its periodic replicas in the direction perpendicular to the surface. The in plane size of the  supercell was 38.32 × 66.38 \AA$^2$ (corresponding to 338 Au atoms per Au(111) layer, 192 (96) Au (Tb) atoms in the TbAu\textsubscript{2} layer  ). The electronic states were expanded using a TZV2P Gaussian basis set~\cite{vandevondele2007gaussian} for  hydrogen  and a DZVP basis set for Au and Tb species. A cutoff of 1600 Ry was used for the plane-wave basis set. Norm-conserving Goedecker–Teter–Hutter pseudo-potentials~\cite{goedecker1996separable} were used to represent the frozen core electrons of the atoms. We used the Perdew–Burke–Ernzerhof parameterization for the generalized gradient approximation of the exchange-correlation functional~\cite{perdew1996generalized}. To account for van der Waals interactions, we used the D3 scheme proposed by Grimme~\cite{grimme2010consistent}. To obtain the equilibrium geometry, we kept the atomic positions of the bottom two layers of the slab fixed to the ideal bulk positions, and all other atoms were relaxed until forces were lower than 0.005~eV \AA$^{-1}$.
The calculations of the DOS and equilibrium geometries presented in Figure 4 of the main manuscript were done with the Quantum ESPRESSO AiiDAlab app~\cite{aiidalab-qe-github-repo} based on the Quantum ESPRESSO code~\cite{Giannozzi2020}. We used the PBE exchange correlation functional with a Hubbard correction (DFT+U~\cite{DFTU2009, DFTU2014} parameter (U) of 6~eV  for Tb 4f electrons. A grid of  90 x 90 k-points was used to sample the Brillouin zone. Pseudopotentials from the SSSP~\cite{SSSP_cite} precision library were used. The simulation cell contained a slab with 7 Au(111) layers and one TbAu\textsubscript{2} layer~\cite{correa2017self}. 

\section{Acknowledgement}\label{acknowledge}
 
This research was supported by, the Swiss National Science Foundation under Grants No. 200020-212875 and TMPFP2-210093, 200020-187617,  SNF-PiMag under grants CRSII5 205987 and the NCCR MARVEL, a National Centre of Competence in Research, funded by the Swiss National Science Foundation (grant number 205602). This
This work was supported by a grant from the Swiss National Supercomputing Centre (CSCS) under project ID s1141.  We acknowledge PRACE for awarding access to the Fenix Infrastructure resources at CSCS, which are partially funded by the European Union’s Horizon 2020 research and innovation program through the ICEI project under the grant agreement No. 800858. We greatly appreciate financial support from the Werner Siemens Foundation (CarboQuant).
J.W.~acknowledges funding by the German Research Foundation (Deutsche Forschungsgemeinschaft, DFG) via the Emmy Noether Programme (project number 503985532), CRC1277 (project number 314695032, subproject A03) and RTG2905 (project number 502572516). J.W.~thanks Ferdinand Evers for helpful discussions. F.X. thanks the Deutsche Forschungsgemeinschaft (DFG) for a Walter-Benjamin Fellowship (Project No. 452269487) and the SNSF Swiss Postdoctoral Fellowship (grant No.210093).

\bibliography{References}

\clearpage

\section{Supporting Info}

\setcounter{figure}{0}
\setcounter{table}{0}
\renewcommand{\thefigure}{SI\arabic{figure}}
\renewcommand{\thetable}{SI\arabic{table}}

\begin{figure}[!htbp]%
\includegraphics[width=\textwidth,]{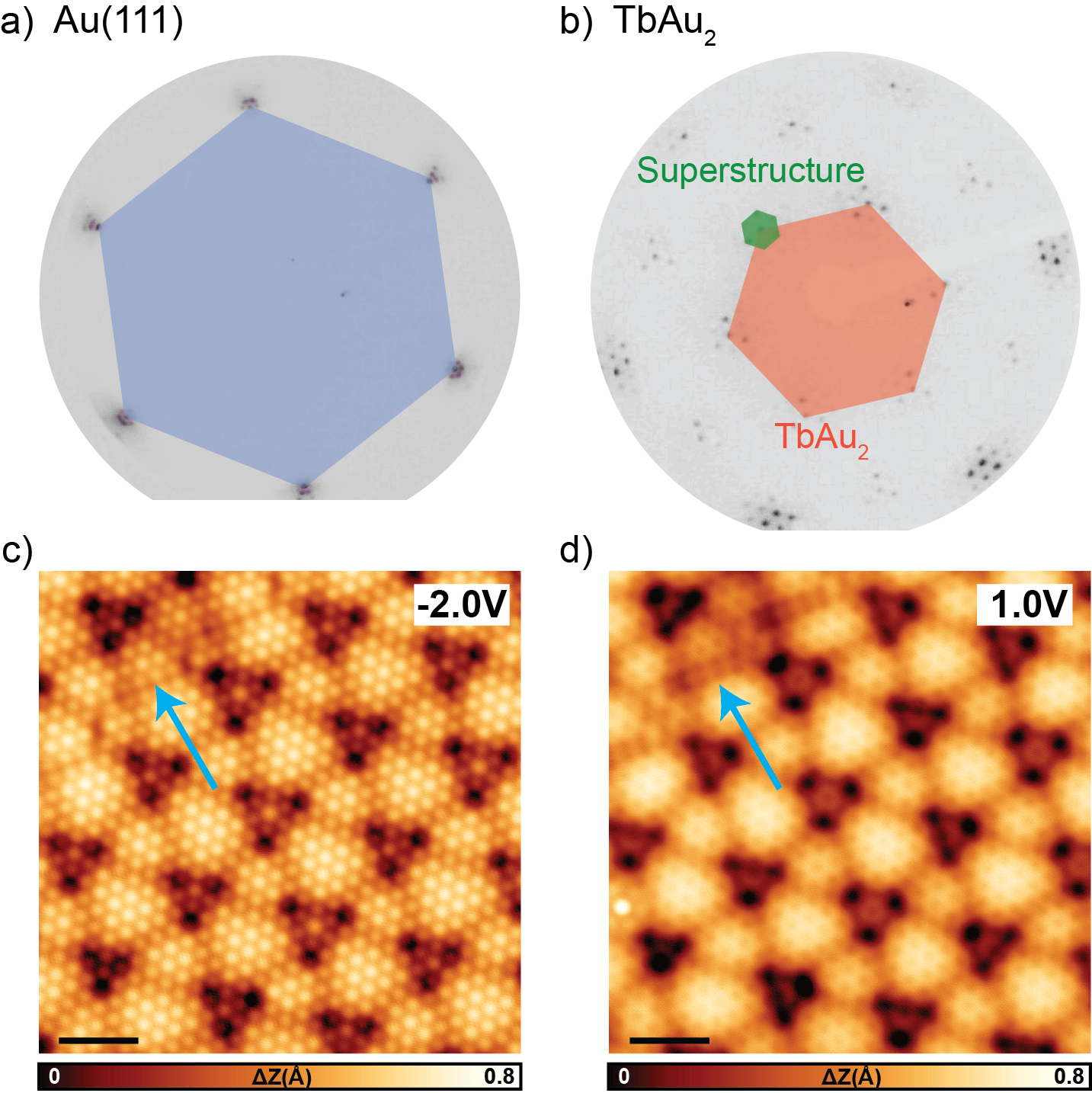}
\caption{\textbf{TbAu\textsubscript{2} surface.}  a) LEED pattern of clean Au(111) at 51~eV, showing the characteristic hexagonal symmetry. b) LEED pattern of the TbAu\textsubscript{2} surface alloy at 60~eV. The hexagonal pattern of TbAu\textsubscript{2} is highlighted in red, while the strain-relief pattern is marked in green. c) Small-scale STM image measured at negative bias voltage, where Tb atoms are imaged as dots (scale bar: 3~nm, scanning parameters: -2~V, 100~pA). d) Small-scale STM image of the same region of panel (c), but acquired at positive voltages. In this case, Tb atoms appear as holes and the Au sublattice is resolved (scale bar: 3~nm, scanning parameters: 1~V, 100~pA). In blue, a grain boundary between two different domains. Along this line, the usual hexagonal periodic structure of the Tb atoms is broken and they form a square lattice (lattice parameter: 0.54~\AA).}
\label{Fig SI1}
\end{figure}

\begin{figure}[!htbp]%
\includegraphics[width=\textwidth,]{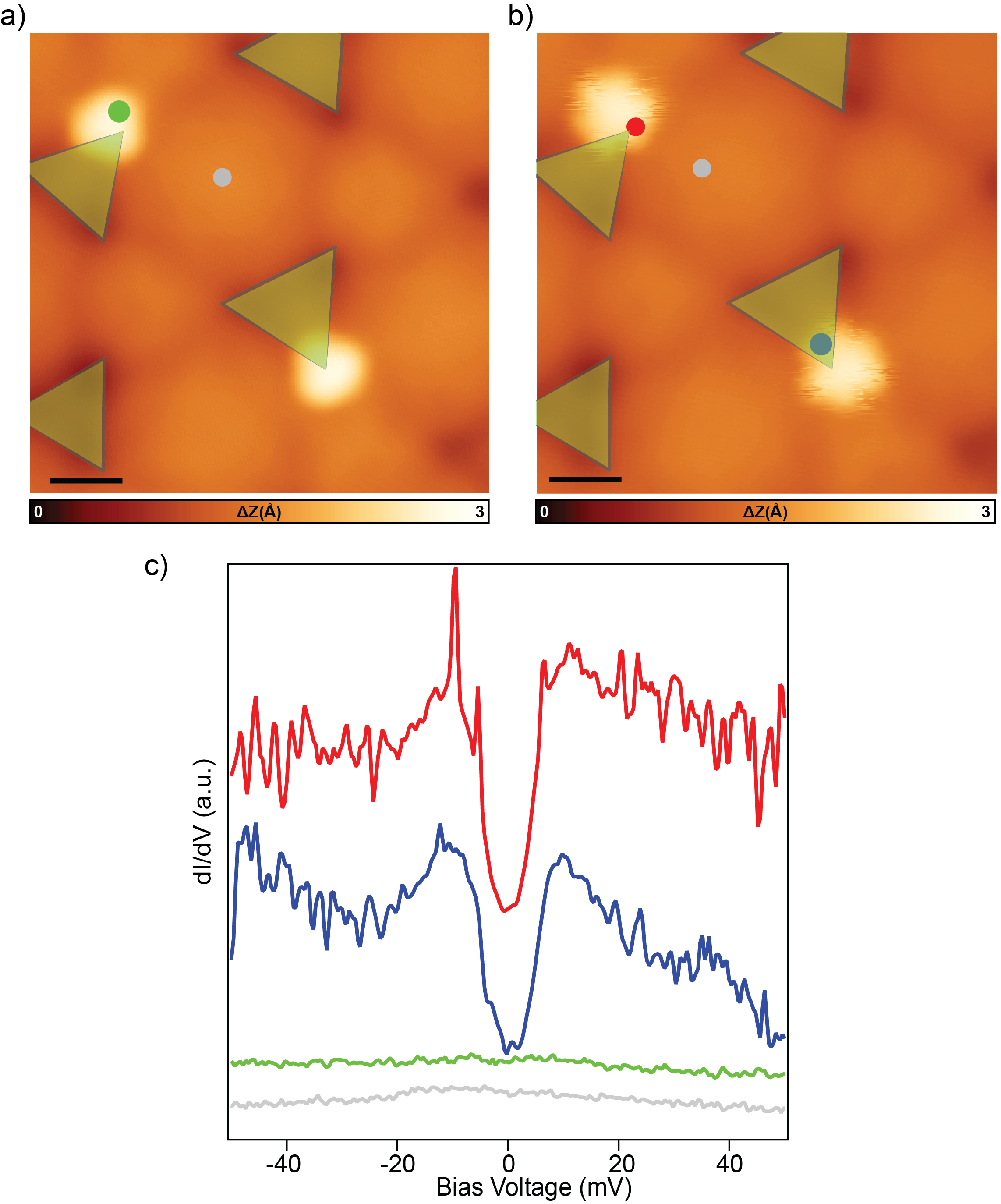}
\caption{\textbf{Complete tip-induced manipulation of 2T}. a) STM image of H-2T on TbAu\textsubscript{2} after sublimation. The molecules exhibit a blunt-triangular shape (Scale bar: 1~nm. scanning parameters: -0.1~V 100~pA). b) STM image of 2T on TbAu\textsubscript{2}. After voltage pulse, the molecules exhibit a hexagonal shape (Scale bar: 1~nm. scanning parameters: -0.1~V 100~pA). Note that after voltage pulse, the molecule are not stable but they flip around their axes c) \didv curves at different molecular states (precise locations are reported in panel a and b). Before tip-induced manipulation, the molecule does not exhibit any relevant features (green spectra). After removal of the extra H atom, the curves show the opening of a gap of 20~mV. (Open feedback parameters: $V$: -100~mV, $I$: 800~pA; V\textsubscript{rms}: 2~mV) }
\label{Fig SI2}
\end{figure}

\begin{figure}[!htbp]%
\includegraphics[width=\textwidth,]{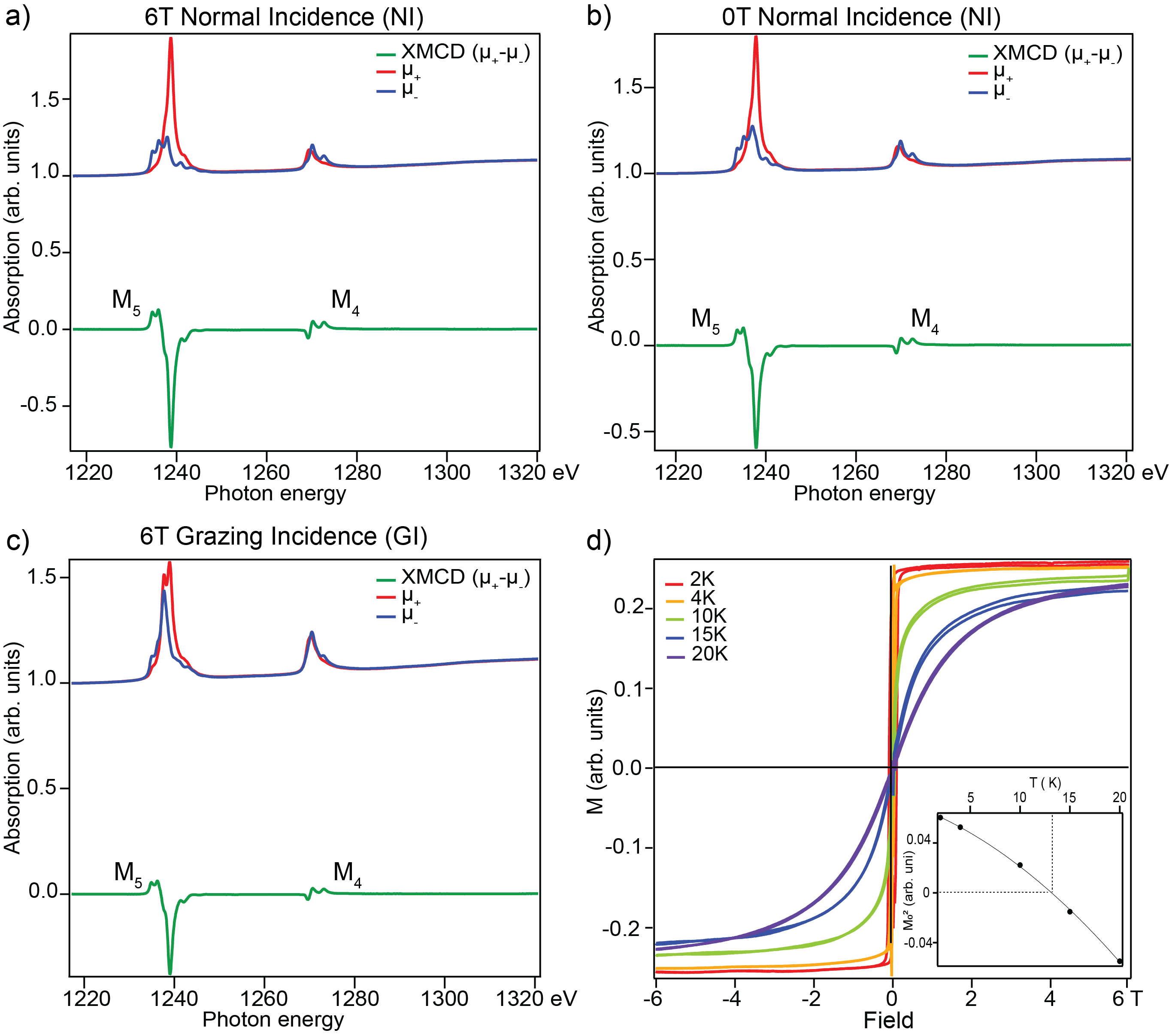}
\caption{\textbf{XAS and XMCD spectra of TbAu\textsubscript{2}}. XAS and XMCD spectra of TbAu\textsubscript{2} measured at the Tb M\textsubscript{4,5} edges at different experimental conditions: a) Normal Incidence (NI) at 6~T, b) Normal Incidence (NI) at 0~T  and c) Grazing Incidence (GI) at 6~T. Absorption spectra with left- and right-circularly polarized light are shown in red and blue, respectively, while the resulting XMCD curves are displayed in green. The presence of a XMCD peak at 0~T under NI confirms the ferromagnetic nature of TbAu\textsubscript{2}. The stronger XMCD signal under normal incidence (NI) indicates a preferred out-of-plane easy-magnetization axis of TbAu\textsubscript{2}. d) Hysteresis curves measured as a function of the temperature in NI. Below 10~K, the curves preserve a step-like shape; above 15~K the shape is more S-like crossing zero. Inset: Arrot plot analysis to extract the T\textsubscript{C} temperature. Similar XAS and XMCD curves were found for other Tb containing compounds, in particular Tb double-decker systems adsorbed on different substrates~\cite{diller2019magnetic,wackerlin2016single}. This confirms the trivalent character of the Tb atoms in TbAu\textsubscript{2}, as common to other rare-earth surface alloys.}
\label{Fig SI3}
\end{figure}

\begin{figure}[!htbp]%
\includegraphics[width=\textwidth,]{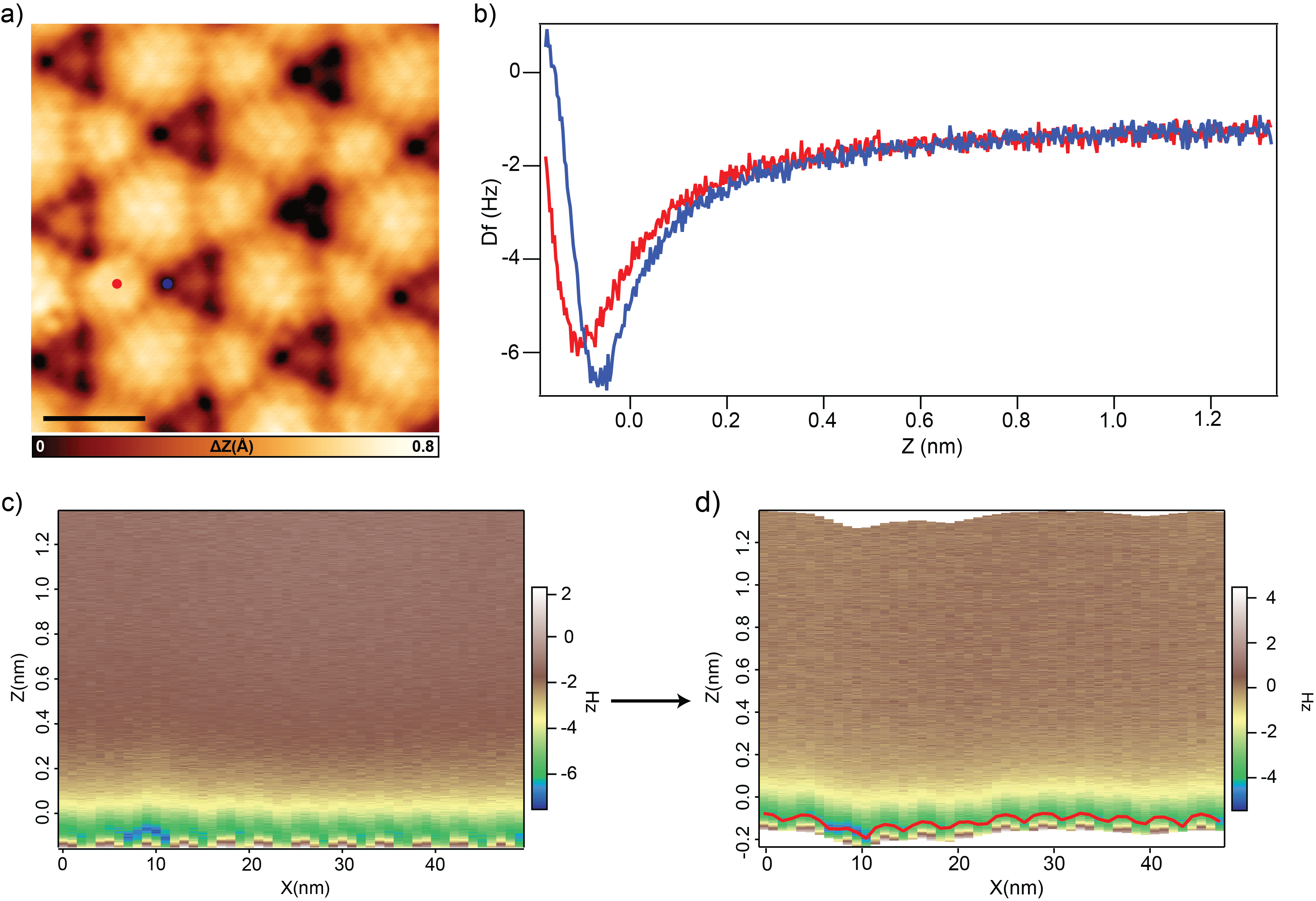}
\caption{\textbf{Height profile variations of TbAu\textsubscript{2}}. a) STM image of TbAu\textsubscript{2} (scanning parameters: -5~mV 200~pA. Scale bar 4~nm). b) Frequency shift ($\Delta f$), as a function of tip-height, at two different points of surface superstructure (panel a). The two curves have different minima corresponding to variations in the atomic heights. c) Frequency shift ($\Delta f$) as a function of tip height along a line crossing the surface superstructure (as shown in \ref{Fig3}a and b). d) Extracted height profiles from the frequency shift data, after vertical offsetting to correct for STM topography variations and subtraction of long-range force background. The red line represents the relative height $\Delta z^\ast$ of the $\Delta f$ minima, as presented in Figure \ref{Fig3}c of the main text.}
\label{Fig SI4}
\end{figure}

\begin{figure}[!htbp]%
\includegraphics[width=\textwidth,]{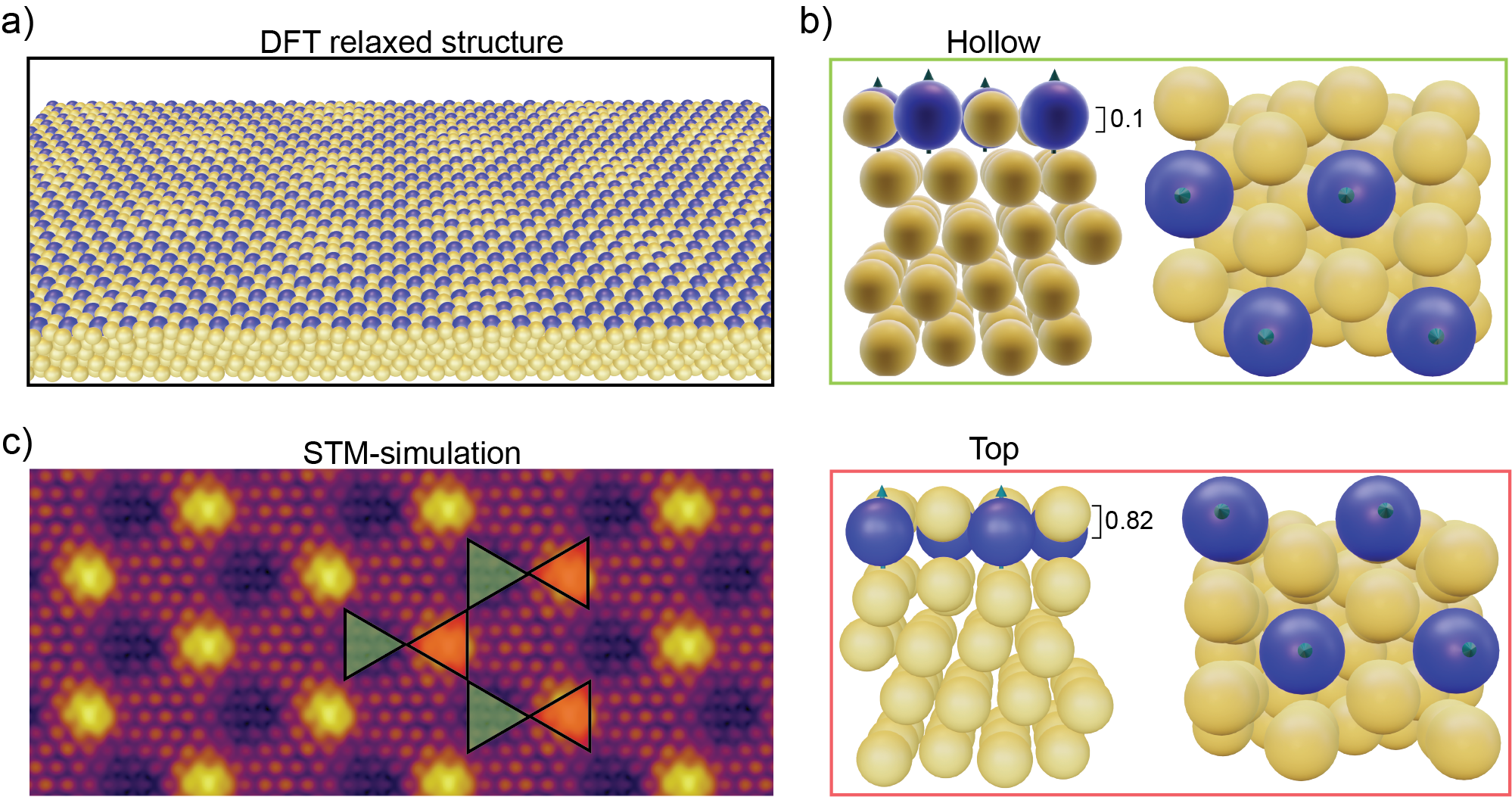}
\caption{\textbf{DFT-relaxed structure of TbAu\textsubscript{2} on Au(111)}. a) DFT-relaxed structure of TbAu\textsubscript{2} on Au(111), with Au atoms in yellow and Tb atoms in blue. The surface exhibits a hexagonal periodic superstructure. b) Vertical displacements analysis for two different  stacking geometries, Hollow (Tb atoms on a hole between three Au atoms) and Top (Tb atom sit directly on top of a Au atom).  In the Top configuration, the Tb atoms are shifted downward relative to neighboring Au atoms, whereas in the Hollow configuration, the Tb atoms are slightly elevated compared to the Au surface plane. Relative displacements (in pm) are indicated in each panel. c) STM simulations of TbAu\textsubscript{2} from panel a. The resulting pattern closely resembles the experimentally observed strain-relief superstructure}
\label{Fig SI5}
\end{figure}

\begin{figure}[!htbp]%
\includegraphics[width=\textwidth,]{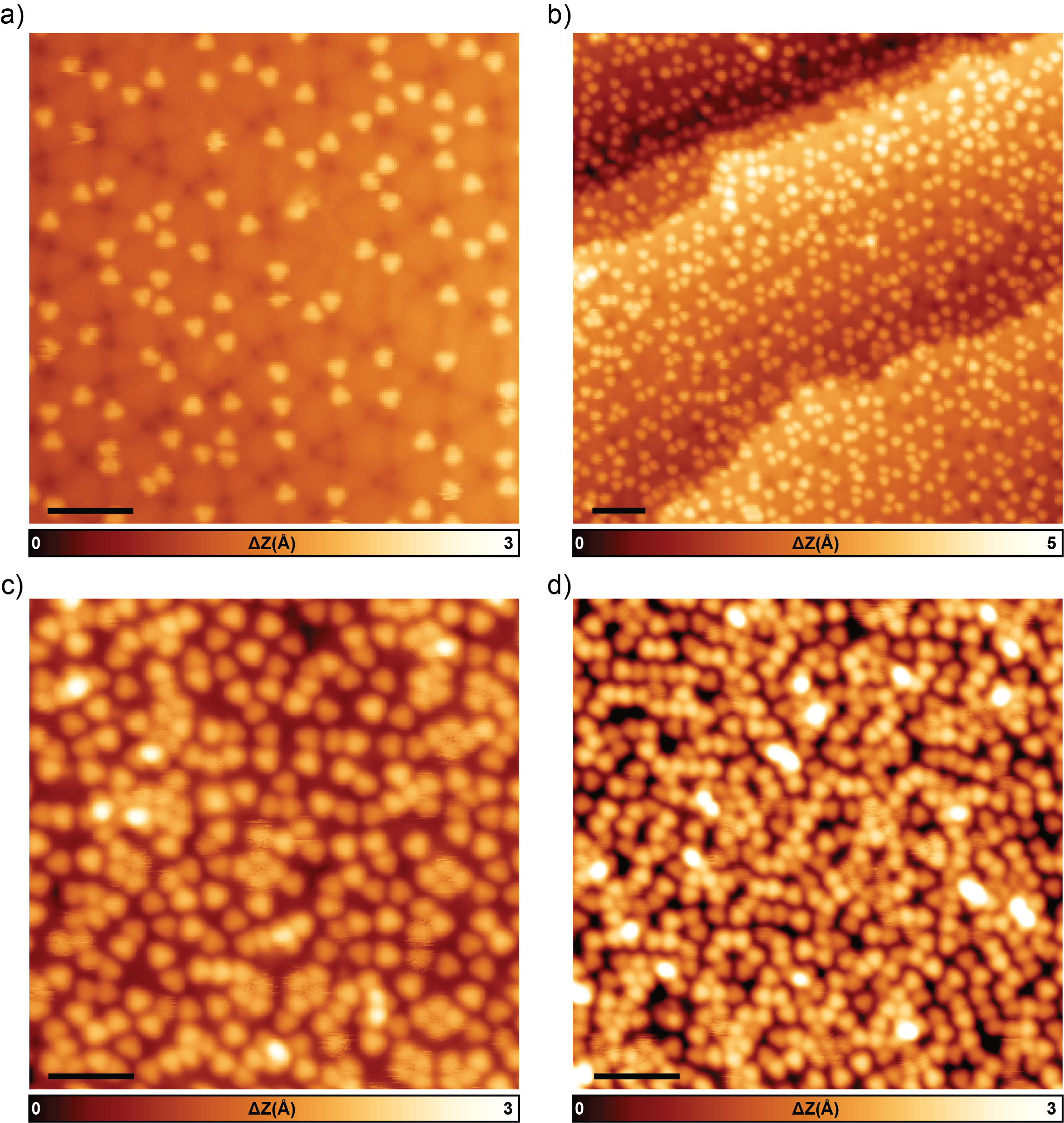}
\caption{\textbf{Coverage increase of H-2T molecules on TbAu\textsubscript{2}}. STM images showing the evolution of H-2T adsorption on TbAu\textsubscript{2} to form 1-ML. Panel b) shows an intermediate stage, roughly 30$\%$, where the molecules are adsorbed occupying the three corners of the Hollow configurations. After all these preferential sites are saturated, H-2T units start to be present in the other regions of the surface superstructure (panel c and d). At this stage, the underlying surface is not distinguishable ( a) scale bar: 4~nm, scanning parameters: $V$:-1~V, $I$: 100~pA. b) scale bar: 5~nm, scanning parameters: $V$: -1~V, $I$: 40~pA. c) scale bar: 4~nm, scanning parameters: $V$: -1~V, $I$: 100~pA. d)  scale bar: 5~nm, scanning parameters: $V$: -1~V, $I$: 100~pA).}
\label{Fig SI6}
\end{figure}

\begin{figure}[!htbp]%
\includegraphics[width=\textwidth,]{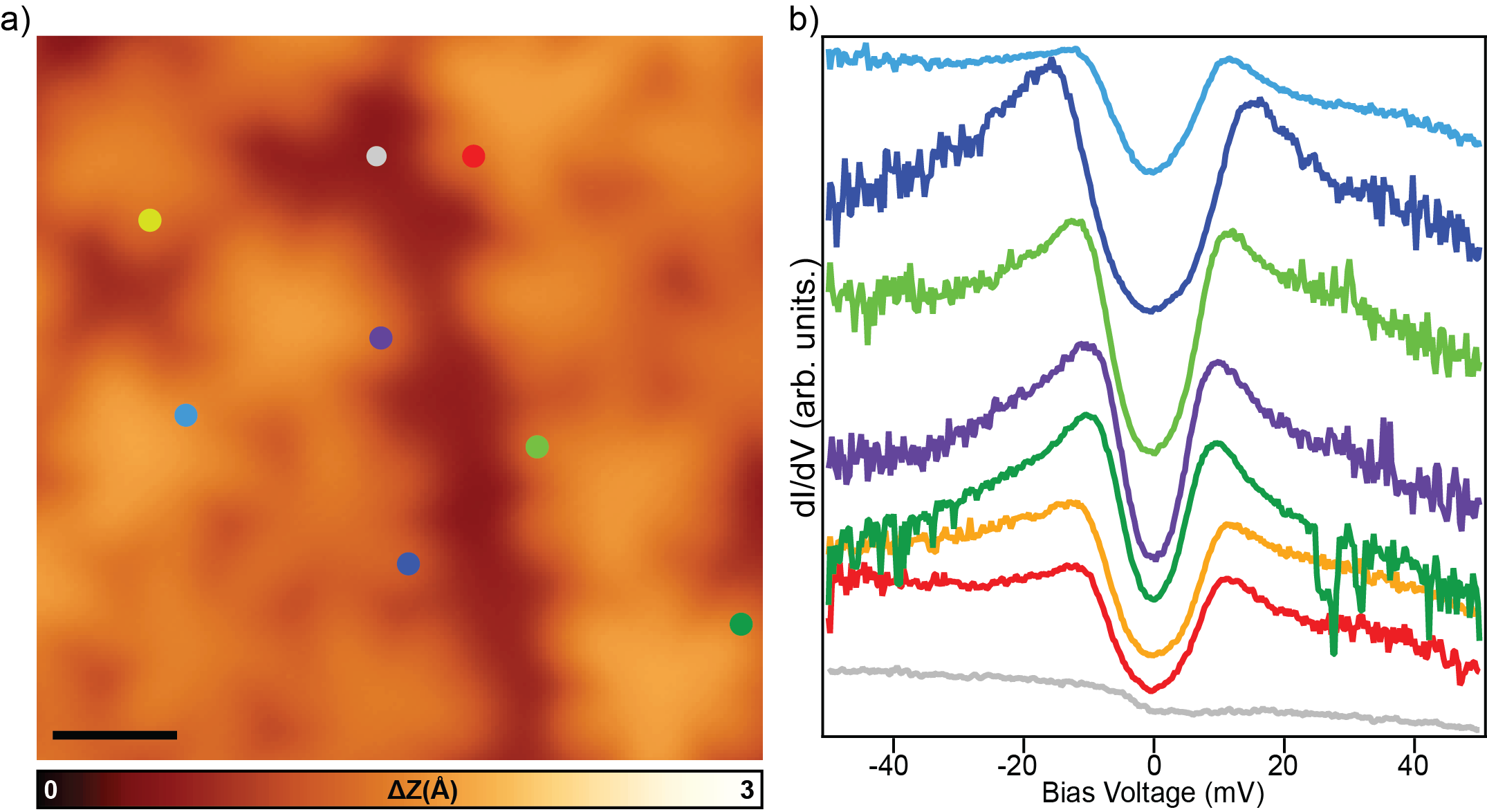}
\caption{\textbf{Low-bias spectroscopy at high coverage on TbAu\textsubscript{2}}. a) STM images of 2T molecules on TbAu\textsubscript{2} at high coverage. The molecules are scattered at different regions of surface superstructure but the surface pattern underneath is not visible (scale bar: 500~pm, scanning parameters: $V$: -50~mV, $I$: 50~pA. ) b) Low bias spectroscopy on different molecules (locations in panel a). All the curves exhibit a split peak, with energy gaps ranging from 22~mV to 30~mV (Open feedback parameters: $V$: -0.05~V, $I$: 700~pA; V\textsubscript{rms}: 2~mV) }
\label{Fig SI7}
\end{figure}

\begin{figure}[!htbp]%
\includegraphics[width=\textwidth,]{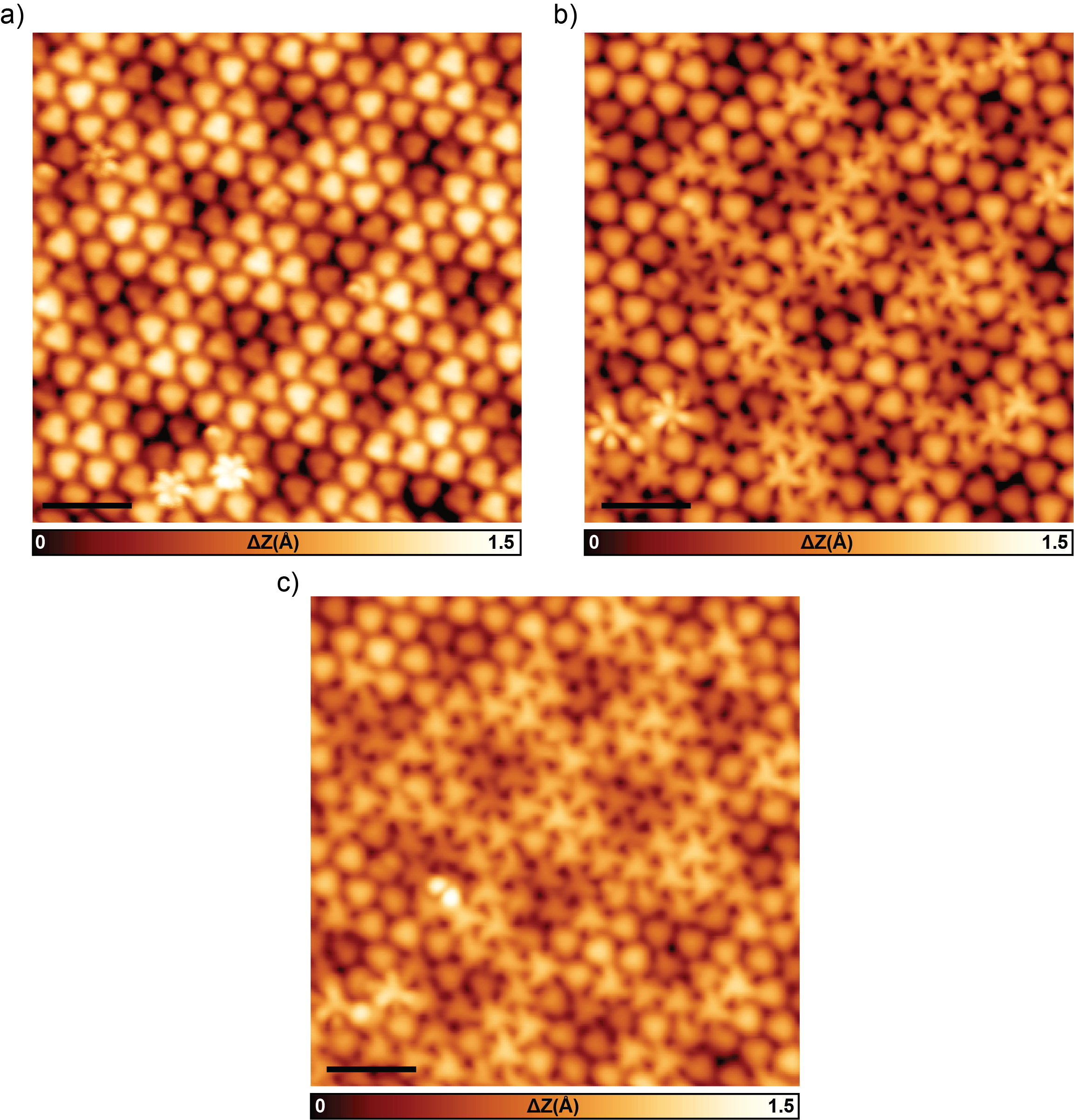}
\caption{\textbf{Progressive manipulation of H-2T molecules}. STM images of the same region of TbAu\textsubscript{2} with progressive tip induced manipulation of the molecules. Due to the high number of molecules, the tip-manipulation was done in multiple steps. a) STM image after sublimation where all the molecules have the H atoms (scale bar: 2~nm, scanning parameters: $V$: -0.15~V,$I$: 100~pA). b) STM image after scanning at 2.8~V and 100~pA. From a large number of molecules, the additional H have been cleaved off (scale bar: 2~nm, scanning parameters: $V$: -0.10~V,$I$: 50~pA). c)  STM image after locally removing the H atoms using the STM tip. Because of the close proximity of the molecules, usually three or four units are manipulated at the same time (scale bar: 2~nm, scanning parameters: $V$: -0.10~V,$I$: 50~pA).}
\label{Fig SI8}
\end{figure}

\begin{figure}[!htbp]%
\includegraphics[width=\textwidth,]{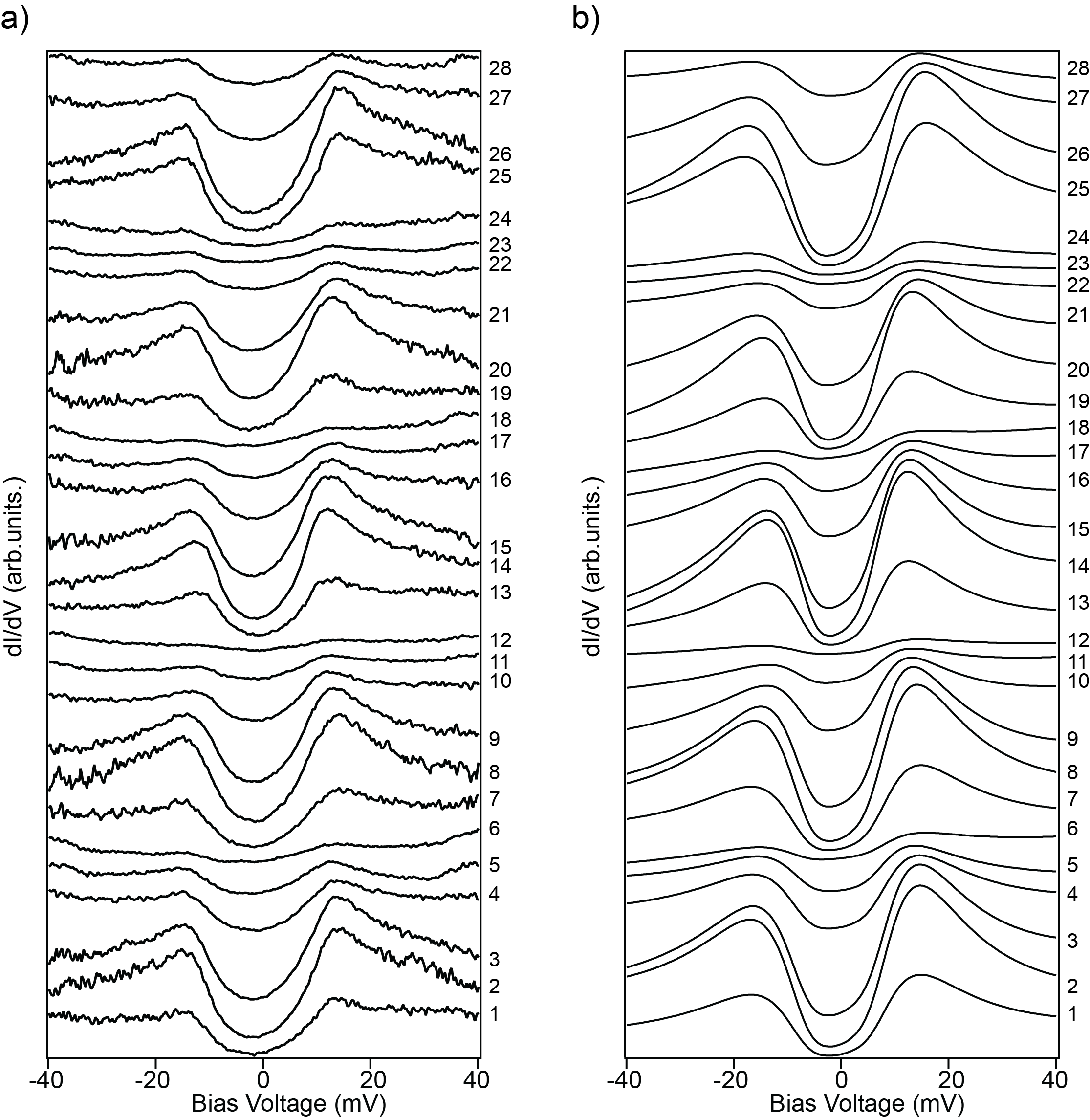}
\caption{\textbf{Spectra taken along a line crossing more unit cells}. a) \didv spectra and b) corresponding NRG fitting taken along a line (white dashed line in Figure 5e). All the spectra are showing a split of the peak with variations depending on the location. (open feedback parameters: $V$: -0.04~V, $I$:900~pA; V\textsubscript{rms}: 1~mV) }
\label{Fig SI9}
\end{figure}

\begin{figure}[!htbp]%
\includegraphics[width=\textwidth,]{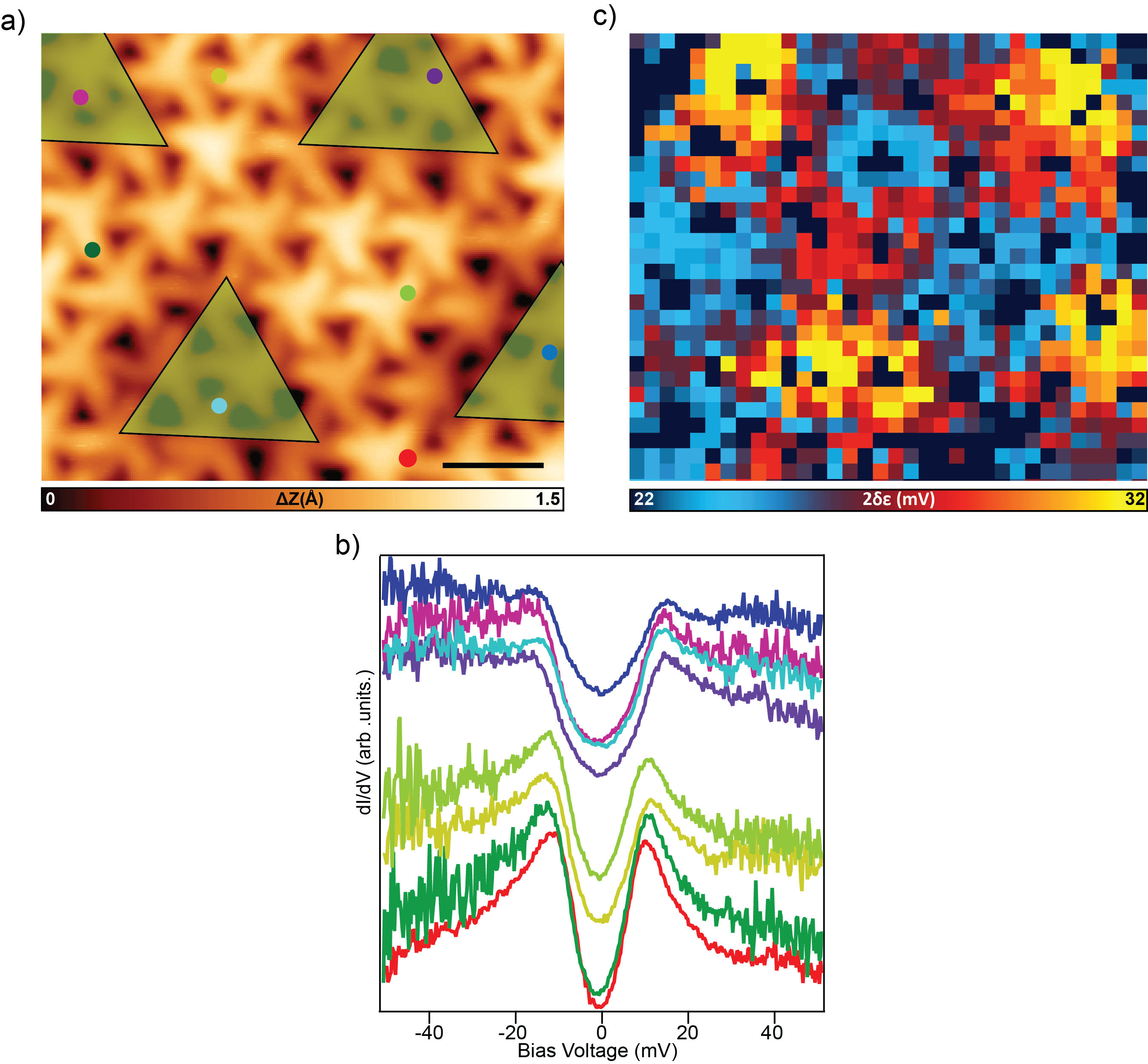}
\caption{\textbf{Exchange interaction maps on large surface}. a) STM image of 2T molecules on TbAu\textsubscript{2} (scale bar: 1~nm, scanning parameters: -0.05~V, 40~pA). b) The spectra exhibit variations in peak splitting depending on the adsorption site. (Open feedback parameters: $V$: -0.05~V,$I$: 900~pA; Vrms: 1~mV). c) Spatial mapping of the effective Zeeman splitting, revealing that the highest values are localized on the dark-triangle regions of the TbAu\textsubscript{2} superstructure.}
\label{Fig SI10}
\end{figure}

\clearpage
\section{Tip-induced manipulation}

The manipulation of the closed-shell H-2T molecules into the open-shell unit 2T was done in situ by using the STM tip. This procedure has been already tested on a number of different molecular systems~\cite{wang2022aza, zhao2024tailoring, turco2023observation}. In particular, the tip was positioned above one of the corner of the molecule and, keeping the current to few tens of pA (20-50 pA), the voltage was progressively increased until a jump in the tip height was detected. we repeated this same procedures over tens of molecules and the activation voltage was around 3.3-3.6~V, depending on the tip apex. 

\clearpage

\end{document}